\documentclass[11pt]{elsarticle}
\usepackage{amsmath,amsfonts, amssymb,graphicx,geometry,color,miller,tcolorbox,soul} 
\usepackage[english]{babel}
\usepackage{blindtext}
\usepackage{isomath}
\usepackage{siunitx}       
\usepackage{tikz}
\newcommand*\circled[1]{\tikz[baseline=(char.base)]{
            \node[shape=circle,draw,inner sep=2pt] (char) {#1};}}
    
\journal{J. Mech. Phys. Solids}

\makeatletter
\def\ps@pprintTitle{%
 \let\@oddhead\@empty
 \let\@evenhead\@empty
 \def\@oddfoot{\centerline{J. Mech. Phys. Solids Accepted Manuscript}}%
 \let\@evenfoot\@oddfoot}
\makeatother

\newtcbox{\mybox}{nobeforeafter,colframe=black,colback=white,boxrule=0.5pt,arc=6pt,
  boxsep=0pt,left=3pt,right=3pt,top=3pt,bottom=4pt,tcbox raise base}

\begin{document}

\begin{frontmatter}

\title{Influences of Granular Constraints and Surface Effects on the Heterogeneity of Elastic, Superelastic, and Plastic Responses of Polycrystalline Shape Memory Alloys}

\author[csmaddress,nuaddress]{Harshad M. Paranjape\corref{correspondingauthor}}
\ead{hparanja@mines.edu}
\cortext[correspondingauthor]{Corresponding Author}
\author[nuaddress]{Partha P. Paul}
\author[apsaddress]{Hemant Sharma}
\author[apsaddress]{Peter Kenesei}
\author[apsaddress]{Jun-Sang Park}
\author[cmsaddress]{T. W. Duerig}
\author[nuaddress]{L. Catherine Brinson}
\author[csmaddress]{Aaron P. Stebner}

\address[csmaddress]{Mechanical Engineering, Colorado School of Mines, Golden, CO 80401, USA}
\address[nuaddress]{Mechanical Engineering, Northwestern University, Evanston, IL 60208, USA}
\address[apsaddress]{Advanced Photon Source, Argonne National Laboratory, Argonne, IL 60439, USA}
\address[cmsaddress]{Confluent Medical Technologies, Fremont, CA 94539, USA}

\begin{abstract}
Deformation heterogeneities at the microstructural length-scale developed in polycrystalline shape memory alloys (SMAs) during superelastic loading are studied using both experiments and simulations. {\it In situ} X-ray diffraction, specifically the far-field high energy diffraction microscopy (ff-HEDM) technique, was used to non-destructively measure the grain-averaged statistics of position, crystal orientation, elastic strain tensor, and volume for hundreds of austenite grains in a superelastically loaded nickel-titanium (NiTi) SMA. These experimental data were also used to create a synthetic microstructure within a finite element model.  The development of intragranular stresses were then simulated during tensile loading of the model using anisotropic elasticity. Driving forces for phase transformation and slip were calculated from these stresses. The grain-average responses of individual austenite crystals examined before and after multiple stress-induced transformation events showed that grains in the specimen interior carry more axial stress than the surface grains as the superelastic response ``shakes down". Examination of the heterogeneity within individual grains showed that regions near grain boundaries exhibit larger stress variation compared to the grain interiors. This intragranular heterogeneity is  more strongly driven by the constraints of neighboring grains than the initial stress state and orientation of the individual grains.
\end{abstract}


\end{frontmatter}

\section{Introduction}
\label{sec:intro}

Shape memory alloys (SMAs) are a class of materials that may exhibit superelasticity, or the ability to fully recover large inelastic deformations induced by mechanical loading. The large inelastic strain during these events arises from a diffusionless solid-solid phase transformation between phases with high and low crystallographic symmetry. Equiatomic, polycrystalline nickel-titanium (NiTi) SMAs in particular can recover strains of up to 6\% in transforming between a cubic austenite (B2) and a monoclinic martensite (B19$^\prime$) phase. Because of this remarkable behavior, they are used for a variety of commercial applications \citep{duerig_overview_1999, mohd_jani_review_2014, otsuka_shape_1999}. Due to the unique properties exhibited by SMAs and the resultant commercial interest, SMAs have received persistent interest from the scientific community. Empirical and theoretical studies have investigated a variety of phenomena related to the stress-induced phase transformation including the crystallography 
\citep{bhattacharya_microstructure_2003, ball_fine_1987, abeyaratne_kinetic_1991, wechsler_theory_1953, bowles_crystallography_1954}, the influence of microstructure and processing on performance \citep{bhattacharya_recoverable_1995, bhattacharya_symmetry_1996, gall_tension-compression_1999, kimiecik_grain_2015, kimiecik_effect_2016, stebner_situ_2015,pelton_situ_2015, schaffer_fatigue_2009,cai_effect_2014}, the inelastic nature of deformation, and the inevitable coupling between phase transformation and plastic deformation 
\citep{bowers_characterization_2014, norfleet_transformation-induced_2009, simon_multiplication_2010, delville_transmission_2011,chowdhury_significance_2016,cai_evolution_2015}. These studies show that, like any deformation process, phase transformation is strongly influenced by microstructural constraint (grain structure, texture) and structural constraint (pores, voids, specimen size, geometry). However, one theme to emerge from these studies is that the constraint can reduce transformation strain magnitude, introduce residual (non-reversible) deformation and can lead to material damage and failure --- all deleterious effects from the perspective of applications \citep{eggeler_structural_2004}.

Microstructural constraint can arise from a variety of compatibility requirements. In polycrystals, grains must maintain compatibility across the grain boundaries. At the interfaces between austenite and martensite phases in SMAs an additional constraint exists due to the necessity to form a low-distortion, low elastic energy interface 
\citep{ball_fine_1987, wechsler_theory_1953}. Such constraint is well-documented to result in localized slip and contribute to poor mechanical behavior --- specifically structural and functional fatigue 
\citep{bowers_characterization_2014,norfleet_transformation-induced_2009,simon_multiplication_2010, perkins_martensitic_1983}. Phase transformation at precipitate-matrix boundaries could be constrained due to coherent or semi-coherent nature of the interface and resultant local stress fields 
 \citep{xie_effect_1990, wang_effect_2015, tirry_linking_2009}. Structural constraint on the other hand refers to the effect of very small specimen sizes \citep{chen_size_2011,manchiraju_pseudoelastic_2012} and other structural features such as porosity \citep{paul_feature_2016,zhao_compression_2005}. Among these constraint effects, the influence of grain boundaries on phase transformation as well as the behavior of grains in the specimen interior vs. free surface have received relatively less attention.

There is a general understanding that similarly oriented grains are theoretically expected to produce similar superelastic transformation strain, but instead they produce a range of strains in real polycrystals, potentially due to the microstructural constraints discussed above \citep{kimiecik_grain_2015, merzouki_coupling_2010, mao_situ_2008}. However, studies of the heterogeneous behavior of similarly oriented grains have been limited to 2D or surface observations, hence the grain deformations have been unconstrained in at least one direction and observations of fully confined grains have been lacking. 3D analyses have typically relied on a modeling component to provide statistics about stress-induced martensite formation in SMA polycrystals \citep{paranjape_texture_2014, gall_role_2000}. While the conclusions from these modeling efforts are general, they rely on idealized, synthetic microstructures and it is challenging to validate those findings empirically. Other efforts have utilized oligocrystalline SMAs to analyze some specific phenomena related to microstructural and structural constraint --- e.g., nucleation of multiple martensite variants at grain triple junctions due to complex stress state vs. single variant at grain boundaries \citep{ueland_grain_2013}. The effect of grain constraint on other inelastic deformation mechanisms e.g., plasticity has been explored both experimentally \citep{sachtleber_experimental_2002, thorning_grain_2005} and analytically \citep{mika_effects_1998}. Accumulated plastic strain and lattice rotations near grain boundaries were observed to deviate from the relatively homogenous deformation states about grain centroids. However, similar to phase transformation, those efforts are either limited to 2D or have investigated idealized microstructures.

Efforts documenting the effect of free surfaces on phase transformation in SMAs at the micron length scale are limited and are primarily based on microwire and micropillar experiments. As a consequence, these results are confined to $<$ \SI{500}{\micro \meter} specimens with a limited number of grains. Findings include a higher fatigue life for oligocrystalline microwires compared to polycrystals \citep{ueland_superelasticity_2012} and a transition from multi-domain martensite microstructure to single domain with a reduction in wire size \citep{ueland_transition_2013}. At an even smaller length scale, transmission electron microscopy (TEM) based studies have documented the occurrence of phase transformation in NiTi nano-pillars \citep{ye_direct_2010}, while suppression of transformation is reported in NiTi thin films with grain size less than 50 nm \citep{waitz_size_2008}. While a combination of techniques have been used in these studies of phenomena related to granular interaction and relaxation at free surfaces of laboratory-produced materials, a desire for a 3D experimental investigation of these phenomena within bulk samples taken from commercially-produced alloys still exists.

The advent of new techniques for non-destructive, {\it in situ} 3D characterization has enabled such a study. High energy diffraction microscopy (HEDM), or 3D X-ray diffraction (3DXRD) techniques, can  non-destructively provide spatially resolved microstructure (grain morphologies, phase, crystal orientation) and deformation (lattice strain tensor) information in bulk specimens during thermo-mechanical loading \citep{bernier_far-field_2011, lienert_high-energy_2011, poulsen_three-dimensional_2001,suter_forward_2006}. These techniques have been utilized to study grain-scale phenomena, e.g., intragranular orientation spread and stress spread developed during elastic and plastic deformation in steel \citep{oddershede_deformation-induced_2015,juul_elastic_2016,winther_grain_2017}, grain rotation and intragranular misorientation evolution in Cu \citep{pokharel_-situ_2015}, change in the volume fractions of the domains in ferroelectric materials \citep{oddershede_quantitative_2015}, twin nucleation in Ti \citep{bieler_situ_2013}, and stress evolution in Ti grains \citep{schuren_new_2015}. Specific to SMAs, 3DXRD technique has been used to probe the grain rotation and grain fragmentation in a CuAlBe SMA during superelastic loading \citep{berveiller_situ_2011} and most recently to image the 3D morphology of a stress-induced transformation interface and the austenite stress field in front of the transformation front in a fine-grained (1 to \SI{5}{\micro \meter}), thin (\SI{100}{\micro \meter}) NiTi wire \citep{sedmak_grain-resolved_2016}.

Here, we use this non-destructive, 3D technique to simultaneously characterize grain-resolved deformation and microstructure during mechanical loading, including the evolution in residual stresses in the grains during cyclic loading, trends in the residual stresses in terms of grain position and orientation, and effect of the residual stresses on subsequent phase transformation. We also use the microstructure information from HEDM to construct a realistic synthetic microstructure for anisotropic, elastic simulations to elucidate two specific phenomena. First, we quantify the deformation heterogeneity in surface vs. interior grains in a superelastically cycled SMA. We propose that the origin of this heterogeneity is from the interaction between grain neighborhoods. Second, we quantify the disparity in intragranular stress state in similarly orientated grains with different neighborhoods. We show that this disparity influences the phase transformation characteristics of the grains. The role of HEDM in our study is to furnish grain-averaged characterization of deformation and orientations. The simulations augment the information at sub-grain scale. An understanding of these phenomena is crucial in designing SMAs that are less prone to structural and functional fatigue. The results from this work advance the general understanding of granular interactions in phase transforming materials.

\section{Materials and Methods}
\label{sec:mat_methods}

\subsection{Material and specimen Preparation}
\label{sec:materials}

The material with a nominal composition of Ti-50.9at.\%Ni was received from Nitinol Devices and Components (NDC) as a bar that was cold drawn 33\% and then creep straightened. The bar was then solution treated at \SI{927}{\degreeCelsius} for 15 min followed by a water quench. This solution treatment, determined by trial, was performed to grow the B2 grain size in the material to \SI{50}{\micro \meter} on average. The austenite finish ($A_{f}$) temperature after the heat treatment is \SI{-1.6}{\degreeCelsius}, resulting in superelastic behavior at room temperature. A cylindrical dogbone specimen with 1 mm gage diameter and 1 mm gage length, as shown in Figure \ref{fig:hedm_setup}(b), was turned from the rod using cylindrical/centerless grinding.

\subsection{Tension Experiment with Far-field High Energy Diffraction Microscopy}
\label{sec:methods_hedm}

An {\it in situ} tension experiment was performed, in which the specimen was loaded for 11 cycles in displacement control to a maximum load of approximately 240 N at a rate of 2 mm/s that resulted in a maximum engineering stress of approximately 300 MPa in each cycle and an effective strain rate of 4.4 $\times$ 10\textsuperscript{-4} s\textsuperscript{-1}. The first 10 cycles were performed to ``shake down'' (i.e., stabilize) the macroscopic stress-strain response of the specimen, and more specifically to stabilize retained martensite and dislocation structures. Using the far field HEDM (ff-HEDM) technique, {\it in situ} measurements of the centroids, volumes, orientations and elastic strain tensors of austenite grains in the gage of the specimen were obtained prior to testing (i.e., before the 1\textsuperscript{st} cycle), and then at several load steps during the 11\textsuperscript{th} cycle: \circled{0} at 22 MPa, \circled{1} at 120 MPa, \circled{2} at 193 MPa, \circled{3} at 260 MPa, \circled{4} at 311 MPa, \circled{5} at 263 MPa, \circled{6} at 175 MPa, \circled{7} at 90 MPa and \circled{8} at 3 MPa. 

\begin{figure}[h]
\centering
\includegraphics[width=5.51in]{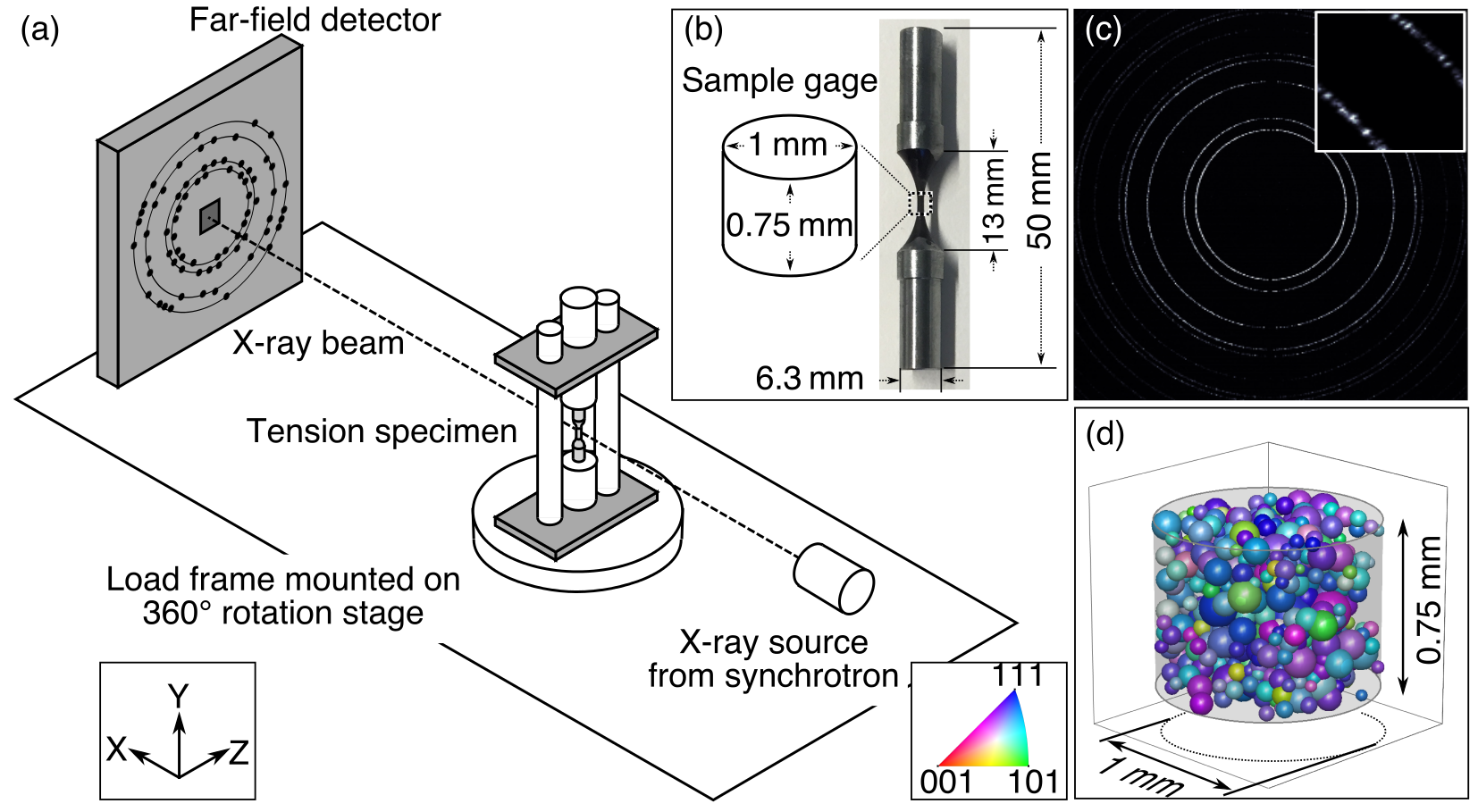}
\caption{A summary of the far-field high energy diffraction microscopy (ff-HEDM) technique. (a) Schematic of a typical ff-HEDM setup to obtain in-situ measurement of grain lattice strains and crystal orientations during an interrupted tension test. (b) A NiTi tension specimen. (c) An area diffraction pattern obtained on a detector placed at approximately \SI{1}{\meter} from the specimen. The spots in the diffraction pattern, shown in the inset, originate from individual grains and contain information about grain position, lattice strain, and crystal orientation. (d) A scatter plot of the grains obtained in the ff-HEDM experiment. The size of the spheres is proportional to the equivalent grain radius and the color uses an inverse pole figure colormap, shown in the inset. Bottom-left inset shows the laboratory coordinate system, which coincides with the specimen coordinate system when the specimen rotation angle ($\omega$) is 0. }
\label{fig:hedm_setup}
\end{figure}

Figure \ref{fig:hedm_setup}(a) shows the experimental setup used in the ff-HEDM experiment. The specimen, shown in Figure \ref{fig:hedm_setup}(b), was mounted in a compact load frame, which itself was placed on a 6-axis goniometer that allowed for alignment and also 360$^{\circ}$ rotation of the specimen about its axis. A 2 mm wide and 0.15 mm tall beam of monochromatic X-rays (71.676 keV) illuminated the specimen gage. A 0.75 mm tall region in the gage center was scanned in 5 layers with the 0.15 mm tall beam at each data point. For each layer, 3600 area detector images were recorded in 0.1$^{\circ}$ intervals on a GE-41RT area detector \citep{lee_synchrotron_2008} placed 759 mm down-stream from the specimen as the load frame was rotated 360$^{\circ}$. The recorded Bragg diffraction angle (2$\theta$) was up to 15$^{\circ}$. A sample diffraction pattern, summed over the 3600 individual images, shows the ``spotty'' nature of the rings in the diffraction pattern in Figure \ref{fig:hedm_setup}(c). Each spot is indicative of a Bragg diffraction condition from a crystal plane within a grain illuminated by the X-ray beam.  Analysis of the spots provided the aforementioned 3D, grain-resolved information. Specifically, the spot intensity furnished grain volume and the spot position furnished grain orientation and grain-averaged lattice strain tensor \citep{sharma_observation_2012, bernier_far-field_2011}. The MIDAS ff-HEDM analysis suite was used to analyze the data \citep{sharma_midas_2016, sharma_fast_2012-1, sharma_fast_2012} and the analysis was performed on the Stampede supercomputer, part of the XSEDE network of computational facilities. The  mean of the austenite lattice parameter measured in the scan prior to the first cycle, $a_{0}$ = 3.0145(3) \AA$\;$ (mean(std. deviation)), was used.  The positional resolution of the ff-HEDM technique is nominally \SI{10}{\micro \meter}, the angular resolution is nominally 0.1$^{\circ}$, and the strain resolution is nominally 10\textsuperscript{-4} \citep{schuren_integrating_2014, sharma_observation_2012, bernier_far-field_2011}. A representative output consisting of grain centroid, position and orientation is shown in Figure \ref{fig:hedm_setup}(d). In the MIDAS analysis, the grains indexed at the start of the 11\textsuperscript{th} cycle are used as seeds for all subsequent analyses. Thus it was possible to track the same grains during loading and unloading.

For this experiment, it was not possible to use MIDAS to analyze the martensitic microstructure with ff-HEDM as the number of martensite crystallites that formed within the austenite grains was too many for the technique (i.e., the martensite data looked like traditional powder diffraction patterns). However, powder diffraction analysis was performed on both phases via summing the data collected at each data point to measure the bulk phase fractions in the gage. The GSAS-II software suite was used for this analysis \citep{toby_gsas-ii_2013}. Note that even though the individual images in the diffraction data show spots and not rings typical of powder diffraction patterns for the austenite phase, summing the signals of  3600 images per layer over all 5 layers results in a powder pattern for the austenite as well.

During the tension experiment, surface strain fields were monitored using digital image correlation (DIC). DIC is a non-contact method of measuring displacements and thus strains by tracking the distortion of a pattern on the surface of the specimen. It has been extensively used to obtain surface strain measurements in SMAs in various types of experiments \citep{bewerse_local_2013, daly_stress-induced_2007, kim_martensite_2011}. To employ this technique, a pattern was created on the specimen surface using black spray paint. Images of the specimen gage surface were taken using a digital camera at approximately 1 s intervals. The images were analyzed using the open-source NCORR software \citep{blaber_ncorr:_2015} and the VIC2D software (Correlated Solutions) to obtain axial strains. From the DIC strain data and load data from the load-frame load cell, global stress-strain curves were constructed. To obtain accurate 2D strain fields on the surface of the 3D cylindrical specimen, a rectangular region of interest centered at the gage in each image was considered. The length of the rectangle is parallel to the specimen loading axis. The breadth of the rectangle is substantially smaller than the length such that the rectangular region can be considered approximately planar.

After the ff-HEDM tension test, the specimen was cleaved length-wise (Y) along the center using wire electrical discharge machining (EDM). One cleaved half was heated to \SI{300}{\degreeCelsius} for 30 min in a vacuum furnace to transform any retained martensite to austenite. After heat treatment, the flat surface of this section was polished and the grain structure and crystal orientations were obtained using electron backscatter diffraction (EBSD) microscopy on a FEI Quanta 600F sFEG ESEM. Tango from Oxford Instruments and open-source MTex software packages \citep{bachmann_texture_2010} were used to perform data processing and visualization.

\subsection{Anisotropic Elastic Simulations Informed by ff-HEDM Experiment}
\label{sec:simulations}

A Voronoi tessellation with a cylindrical domain was created from the B2 grain centroid and orientation data measured from ff-HEDM at the start of 11\textsuperscript{th} cycle (scan \circled{0}). A finite element mesh with $1.48 \times 10^{6}$ elements was created from the tessellation. Hexahedral, full-integration elements  (C3D8) were used in Abaqus to perform the finite element simulations \citep{simulia_abaqus_2008}. The center of the base of the cylindrical model was pinned to suppress rigid body motion and other nodes on the base were assigned in-plane sliding boundary conditions. All nodes on the top surface of the model were constrained in the loading direction to a reference node, but they were allowed to displace in-plane. cubic single crystal elastic properties of $C_{11}$ = 130 GPa, $C_{12}$ = 98 GPa and $C_{44}$ = 21 GPa were used \citep{brill_elastic_1991}. An anisotropic, elastic constitutive law was implemented in an Abaqus user material subroutine (UMAT). Inelastic deformations were not simulated, but the likelihoods of activity of transformation and slip mechanisms were calculated from the stresses at each time step. Using the UMAT, isothermal, elastic loading was simulated to a macro tensile strain of 1\% in 10 increments by applying a displacement boundary condition to the reference node. The UMAT algorithm follows:
\begin{enumerate}
\item Stiffness tensor ($\tensorsym{C}^{\textrm{crystal}}$) for each integration point is transformed to the specimen coordinate system using the crystal orientation of the grain to which the element belongs.
\begin{equation}
\tensorsym{C} = \tensorsym{G}\tensorsym{G}\tensorsym{C}^{\textrm{crystal}}\tensorsym{G^{\textrm{T}}}\tensorsym{G^{\textrm{T}}}.
\end{equation}
Here $\tensorsym{G}$ is the rotation matrix representing the crystal orientation of the element. All subsequent calculations are performed in the specimen frame.

\item Cauchy stress ($\tensorsym{\sigma}$) is calculated from the deformation gradient ($\tensorsym{F}$) given by Abaqus using Hooke's law. 
\begin{subequations}
\begin{align}
        \tensorsym{E} &= \frac{1}{2}(\tensorsym{F}^{T}\tensorsym{F} - \tensorsym{I}), \\
        \tensorsym{T} &= \tensorsym{C} \tensorsym{E}, \\
        \tensorsym{\sigma} &= \frac{1}{\textrm{det} \tensorsym{F}} \tensorsym{F} \tensorsym{T} \tensorsym{F}^{T}.
\end{align}
\end{subequations}
Here $\tensorsym{E}$ is the Green strain and $\tensorsym{T}$ is the 2\textsuperscript{nd} Piola-Kirchhoff stress. The Cauchy stress is returned to Abaqus.

\item Martensite habit plane variant (HPV) plane normal ($\vectorsym{m_{i}}$) and shear ($\vectorsym{b_{i}}$) are calculated using the well-established crystallographic theory of martensite (CTM) \citep{bhattacharya_microstructure_2003}. 
For this calculation, a B2 lattice constant of $a_{0} = 3.0145$ \AA$\;$  from the ff-HEDM measurement and B19$^\prime$ lattice constants of $a$ = 2.889 \AA, $b$ = 4.12 \AA, $c$ = 4.622 \AA $\;$ and $\beta$ = 96.8$^{\circ}$ from \citet{bhattacharya_microstructure_2003} are used. For the B2 $\rightarrow$ B19$^\prime$ phase transformation, there are 192 possible solutions and hence 1 $\leqslant i \leqslant$ 192. Using the parent grain orientation ($\tensorsym{G}$), vectors $\vectorsym{b_{i}^{\textrm{crystal}}}$ and $\vectorsym{m_{i}^{\textrm{crystal}}}$ are transformed to the specimen coordinate system ($\vectorsym{b_{i}} = \tensorsym{G} \vectorsym{b_{i}^{\textrm{crystal}}}$, $\vectorsym{m_{i}} = \tensorsym{G} \vectorsym{m_{i}^{\textrm{crystal}}}$),.

\item Using the Cauchy stress and the habit plane elements ($\vectorsym{b_{i}}$, $\vectorsym{m_{i}}$), the HPV most likely to form is determined based on the maximum work criterion. This HPV is denoted by $P_{t}$.
\begin{equation}
P_{t} \ni \tensorsym{\sigma} \cdot (\vectorsym{b_{t}} \otimes \vectorsym{m_{t}}) = \textrm{max} \{\tensorsym{\sigma} \cdot (\vectorsym{b_{i}} \otimes \vectorsym{m_{i}})\}; \quad 1 \leqslant i \leqslant 192.
\label{eq:max_work_plate}
\end{equation}
For the HPV $P_{t}$, the transformation strain ($\vectorsym{\epsilon_{t}^{\textrm{m}}}$) is calculated using: $\vectorsym{\epsilon_{t}^{\textrm{m}}} = 1/2(\vectorsym{b_{t}} \otimes \vectorsym{m_{t}} + \vectorsym{m_{t}} \otimes \vectorsym{b_{t}})$.

\item The resolved shear stress ($\tensorsym{\sigma_{i}^{\textrm{RSS}}}$) on 12 B2 slip systems is calculated using: $\tensorsym{\sigma_{i}^{\textrm{RSS}}} = \tensorsym{\sigma} \cdot (\vectorsym{b_{i}^{\textrm{slip}}} \otimes \vectorsym{m_{i}^{\textrm{slip}}})$. Here $\vectorsym{b_{i}^{\textrm{slip}}}$ is slip direction and $\vectorsym{m_{i}^{\textrm{slip}}}$ is the slip plane normal. Six B2 slip systems of \hkl{1 1 0}//\hkl<1 0 0> type and six of \hkl{1 0 0}//\hkl<1 0 0> type are considered. The slip system with the maximum resolved shear stress is determined and stored. This slip system is denoted by $S_{t}$.
\end{enumerate}

\section{Results}
\label{sec:results}

\subsection{Far-field High Energy Diffraction Microscopy Results}
\label{sec:results_hedm}

\begin{figure}
\centering
\includegraphics[width=5.51in]{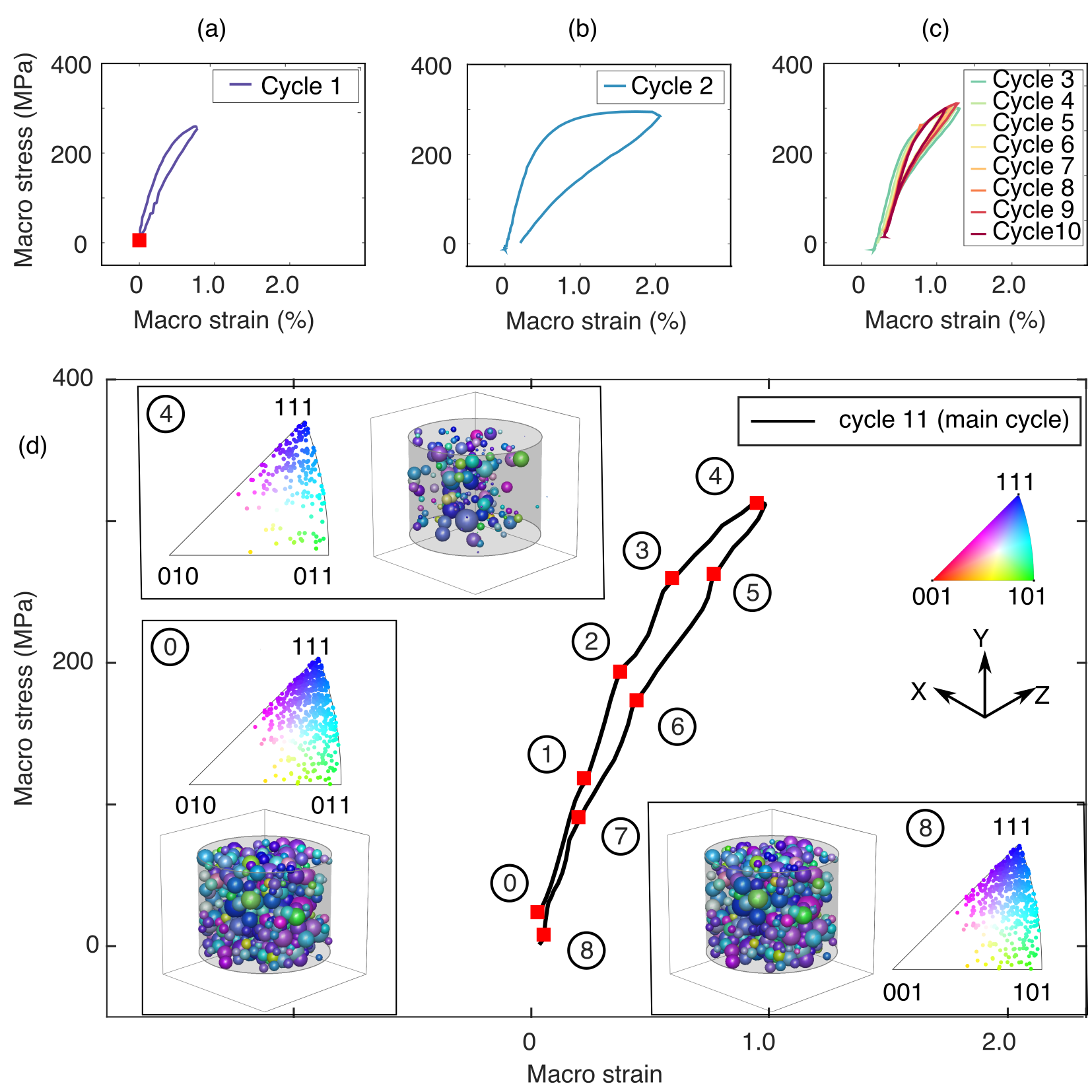}
\caption{Macro stress-strain curves for the 11-cycle tension test. First ten tension cycles are performed to stabilize inelastic deformation. (a) First cycle. An initial ff-HEDM scan is performed before the first cycle, shown by a red square. (b) Second cycle showing transformation plateau. (c) Cycles 3-10 show minimal additional irrecoverable strain accumulation. (d) Main cycle during which 9 ff-HEDM scans are performed. Initial strain is reset to 0. Inverse pole figure and a 3D view of the grain center of mass is shown at three key stages: 0 load (\protect\circled{0}), peak load (\protect\circled{4}) showing fewer B2 grains remaining due to phase transformation, and full unload (\protect\circled{8}) showing near-complete reverse transformation to B2. The grains are colored according to an inverse pole figure colormap.}
\label{fig:ss_macro}
\end{figure}

Our goal in these experiments was to study the micromechanics related to the initiation of transformation. As such, we chose our loading paths to initiate, but not saturate the phase transformation. Figure \ref{fig:ss_macro} shows a summary of the macroscopic stress-strain evolution during the 11-cycle tension test. Cycles 1 to 10 show an evolution in the response. In particular, cycle 2, where the maximum strain is higher than other cycles, shows a residual strain accumulation of 0.22\%. The initiation response of the specimen further stabilizes in cycles 3 to 10. 

The 11\textsuperscript{th} cycle in Figure \ref{fig:ss_macro}(d) shows a non-linear response with hysteresis and full strain recovery. ff-HEDM scan points are labeled \circled{0} to \circled{8}. The non-linearity is more noticeable between scans \circled{2} and \circled{7}. Left-bottom inset shows the B2 grain positions, size, and orientation measurements from ff-HEDM at scan \circled{0}. The material has a strong \hkl<1 1 1> texture with a \hkl<1 1 0> component along the loading direction --- typical of drawn NiTi rods. The left-top inset shows the tracked grain structure and orientations at \circled{4}. A fraction of the grains have transformed to martensite, hence they disappear in the reconstructed data. At full unload, \circled{8}, most of the grains reappear as the material mostly transformed back to austenite.

\begin{figure}
\centering
\includegraphics[width=5.51in]{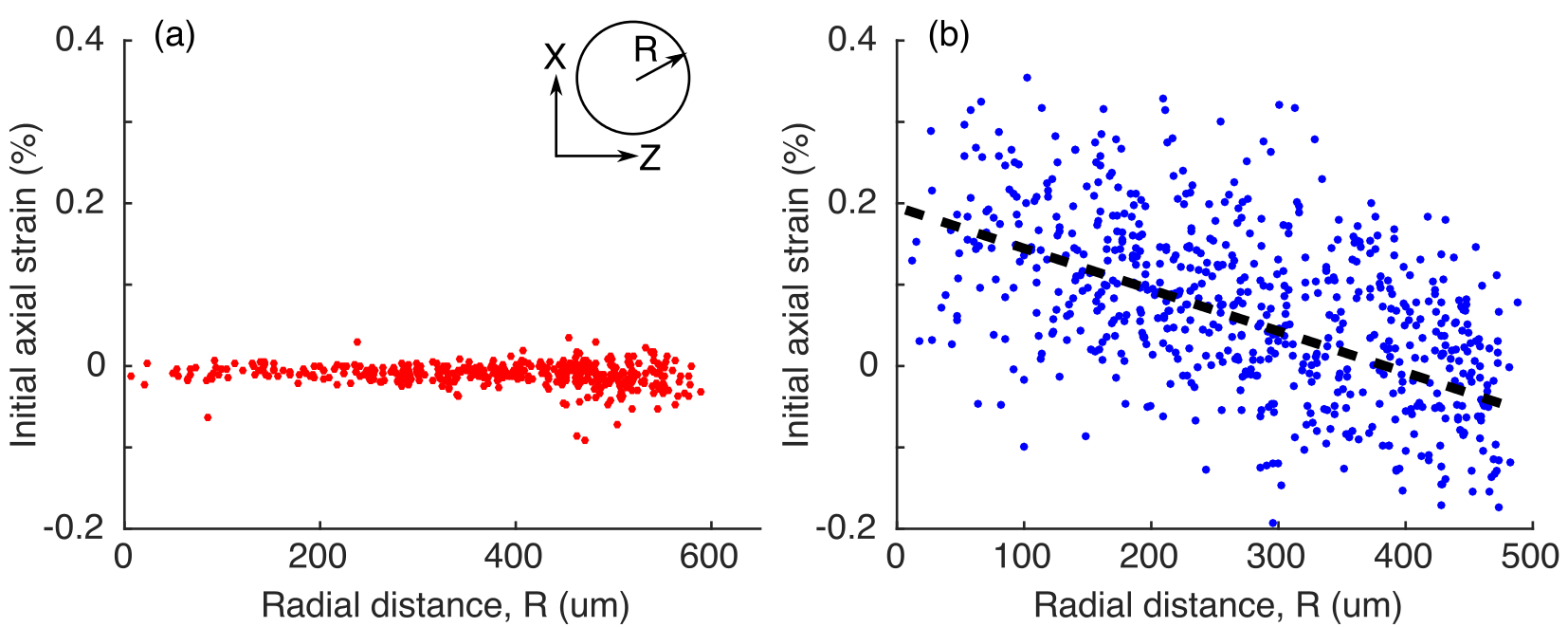}
\caption{Axial lattice strains in the individual grains as a function of the grain radial distance from the specimen center axis. (a) Cycle 1 with nearly-zero axial strains, expected in a virgin material. The inset shows the radial distance in a top-down view of the specimen. (b) Load step \protect\circled{0} in Cycle 11. The grains near the center are in a tensile state vs. grains near the surface which are in weak tension to compression. Note that measurement \protect\circled{0} was conducted at a small tensile stress (22 MPa). Thus the strains are biased in tension. The dotted line shows a linear fit.}
\label{fig:lattice_strains_radial}
\end{figure}

The specimen had axial lattice strains close to zero prior to cycling. Figure \ref{fig:lattice_strains_radial}(a) shows the distribution of axial lattice strain vs. grain centroid distance from the specimen axis at the start of cycle 1. Lattice strains are computed from the distorted and reference lattice parameters for the material and hence represent the elastic component of deformation. The zero strains represent an absence of internal stresses in the virgin material. A heterogeneous strain distribution is observed after cycles 1-10. Figure \ref{fig:lattice_strains_radial}(b) shows a similar plot at scan \circled{0} in the 11\textsuperscript{th} cycle. The lattice strains are the largest for the interior grains and  they decrease with increasing radial distance, albeit with a scatter.

\begin{figure}
\centering
\includegraphics[width=3.54in]{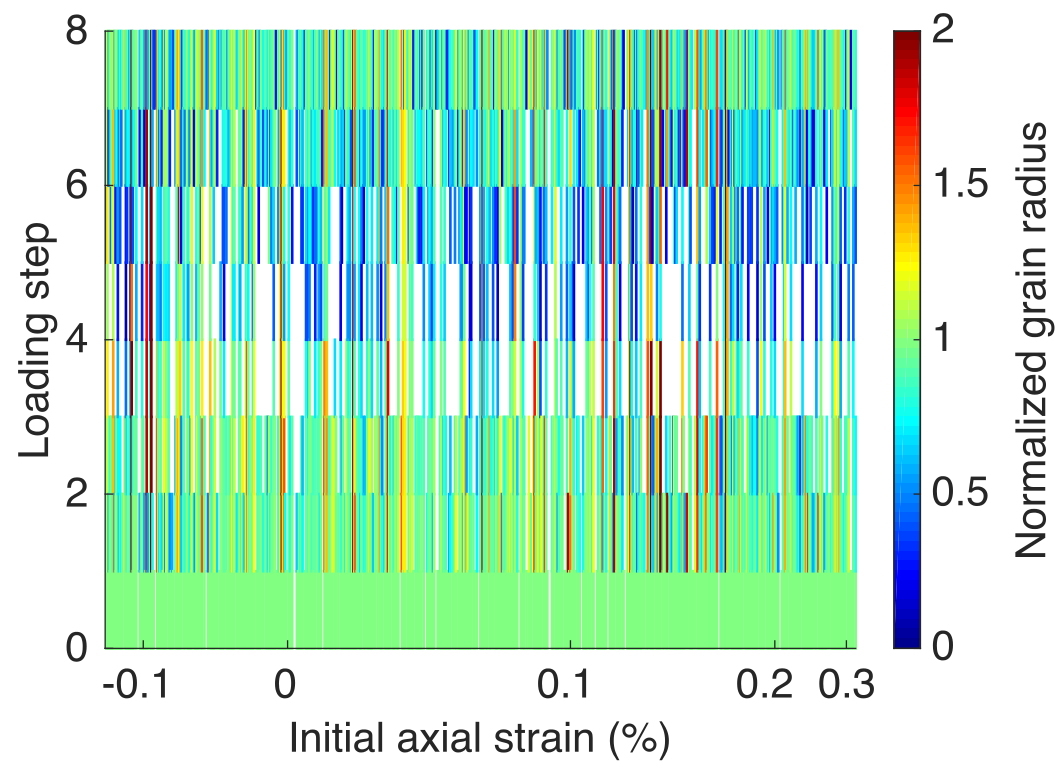}
\caption{Relation between grain tracking during ff-HEDM scans and the axial lattice strain at \protect\circled{0} in Cycle 11. One grain occupies a bin on the X axis. Thus a vertical column represents the history of a grain in terms of its radius. A grain is not tracked when either the  diffraction spots corresponding to the grain become significantly distorted or the grain substantially transforms to B19$^\prime$. Disappearance of a grain is seen as a white-space in the vertical column. A larger number of white spaces at a loading step implies that a larger fraction of grains disappeared from tracking. Loading steps \protect\circled{4} and \protect\circled{5}, which correspond to the peak load, show a larger number of white spaces compared to other loading steps, since a larger fraction of grains are expected to transform to B19$^\prime$ and disappear at these steps. Grains with a large compressive axial strain do not disappear as much as the grains initially in tension. Thus, there are fewer white spaces in the extreme left region of the plot compared to the region on the right. The colors are based on the grain radius normalized with the radius at \protect\circled{0}. In general, the grain radius decreases near peak loads at \protect\circled{4} and \protect\circled{5}, and thus the color changes to blue and cyan. An alternate representation of this data, along with additional annotations, is shown in the Supplementary Data Figure S1.}
\label{fig:grain_tracking}
\end{figure}

Figure \ref{fig:grain_tracking} shows a visualization of 651 tracked grains in cycle 11 as a function of the lattice strain in the grains at \circled{0} in cycle 11. The binning on the X axis is performed such that a single grain is present in a bin. Thus a vertical column shows the evolution in the radius of a single grain. An absence of a vertical bar (i.e., presence of white space) means that the grain transformed to martensite, at least to a size (radius) smaller than the detectable limit of the ff-HEDM setup and analysis employed here (approximately \SI{10}{\micro \meter} radius). Several grains with a tensile initial lattice strain disappear between loading steps \circled{3} and \circled{6}. The grains with a compressive initial axial stress however, do not disappear as much as the tensile grains. The grain radius for all grains, shown using color in the figure, in general, tends to decrease gradually between \circled{0} and \circled{4} and then increases back up. Thus the color of the vertical bars shifts to cyan and blue at steps \circled{4}, \circled{5}. Many grains however, do not transform back to their initial volume (i.e., normalized grain radius not equal to 1). In fact, some grains are larger, while others are smaller at step \circled{8} than their initial size at step \circled{0}. The {\em equivalent} grain radius in the ff-HEDM measurement is obtained from the grain volume assuming a spherical shape. The grain volume itself is determined from the relative intensity of the spots corresponding to the grain and the integrated intensity of the ring to which the spots belong \citep{sharma_observation_2012}. Hence, based on the intensity threshold chosen to identify spots, a certain variation in the measured grain volume is possible. An alternate representation of the data in Figure \ref{fig:grain_tracking} is shown in the Supplementary Data Figure S1. A more quantitative analysis of the indexed grain number and mean grain radius shows a relation to the progress of phase transformation.

\begin{figure}
\centering
\includegraphics[width=5.51in]{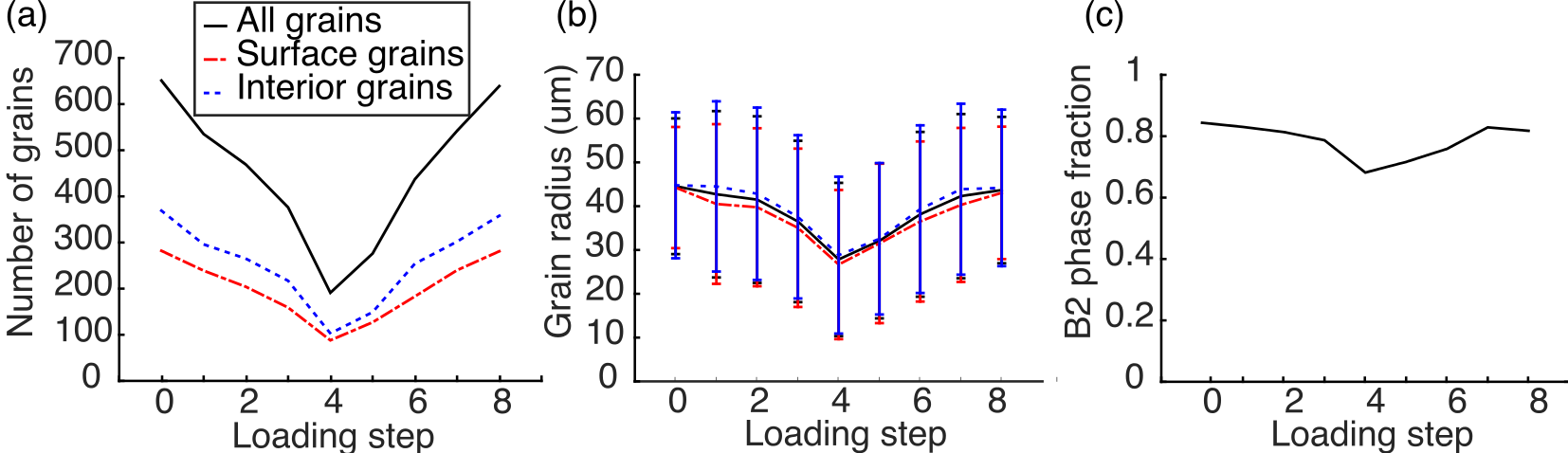}
\caption{Evolution of the number of indexed B2 grains, size, and B2 phase fraction. (a) Number of tracked grains decreases near peak load (step 4) due to phase transformation. (b) Average grain radius decreases during loading (steps 0-4) and increases during unloading (4-8) due to partial phase transformation in the grains. (c) B2 phase fraction from powder analysis during the Cycle 11 decreases during loading and increases during unloading. In (a, b), grains with the centroid more than 0.3 mm away from the specimen axis are classified as surface grains.}
\label{fig:grainnum}
\end{figure}

Figure \ref{fig:grainnum}(a) shows the change in the number of grains indexed during the 11\textsuperscript{th} cycle. At \circled{0}, 651 grains are indexed in the gage. The number decreases to 191 at \circled{4} and then recovers to 639 at \circled{8}. The grains with the centroid less than 0.3 mm from the gage axis (Y axis) are labeled as {\em interior} grains and the remainder are labeled as {\em surface} grains. Based on this criterion, the number of surface vs. interior grains decreases at approximately the same rate during initial part of loading. Between scans \circled{3} and \circled{4} however, interior grains transform at a faster rate.

Figure \ref{fig:grainnum}(b) shows the evolution in mean and standard deviation of grain radii during loading. The initial radius of 53$\pm$\SI{18}{\micro \meter} decreases to 33$\pm$\SI{21}{\micro \meter} at peak load and recovers to 52$\pm$\SI{20}{\micro \meter} on unload. The change in mean grain radius for surface vs. interior grains is similar. The error bars for the radius plots show the standard deviation, which is approximately \SI{20}{\micro \meter} at each loading step.

Figure \ref{fig:grainnum}(c) shows the evolution of B2 phase fraction in the main (11\textsuperscript{th}) cycle, as calculated using Rietveld refinement on the sum of all diffraction patterns at each load step (see Section \ref{sec:methods_hedm}). The initial and final B2 volume fraction is 0.84 and 0.82 respectively. This data reveals that some martensite was retained due to the residual stress that accumulated during the ``shake down'' cycles.  However, little additional martensite was retained in this cycle, and the phase transformation was mostly reversible upon unload. The minimum B2 phase fraction of 0.68 underscores that the transformation was partial at peak load. This result is qualitatively similar to the grain-scale results in (a, b), both of which also reflect partial phase transformation at peak load.

\begin{figure}
\centering
\includegraphics[width=\textwidth]{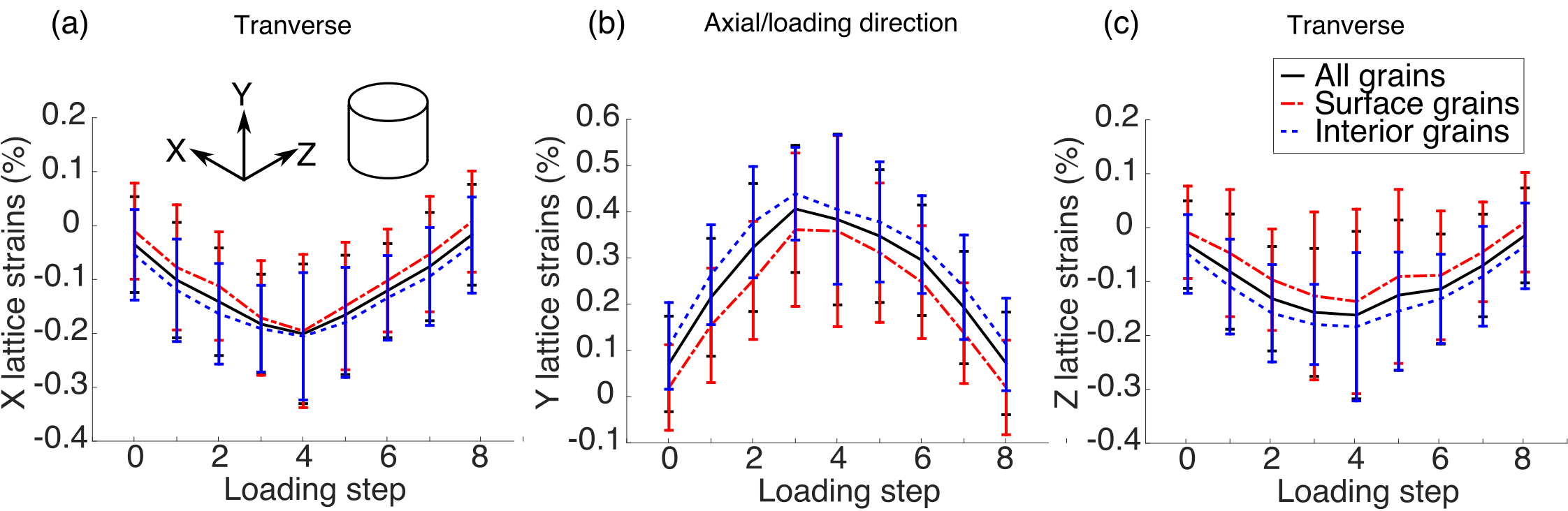}
\caption{Evolution of the lattice strains along specimen X direction (a), Y or axial direction (b) and Z direction (c). Grains with the centroid more than 0.3 mm away from the specimen axis are classified as surface grains. The surface grains, on average, show lower axial strain, but higher strains in the transverse directions.}
\label{fig:lattice_strains}
\end{figure}

Figure \ref{fig:lattice_strains} shows the evolution of lattice strains averaged over the grains in the gage. The strains along the loading (Y) direction in (b) show two trends. First, the mean axial lattice strain in the surface grains is consistently lower than the interior grains. Second, the standard deviation in the strains is large, indicating a highly heterogeneous elastic strain state. The results in the transverse directions, shown in (a, c) show corresponding trends resulting from the Poisson effect. In the case of all three strain components, the strains at \circled{0} and \circled{8} are similar. The mean axial strain at \circled{0} is greater than zero. This is because the specimen was at approximately 22 MPa stress at the start of the 11\textsuperscript{th} cycle.

\subsection{Anisotropic Elastic Simulation Results}
\label{sec:results_sim}

\begin{figure}
\centering
\includegraphics[width=5.51in]{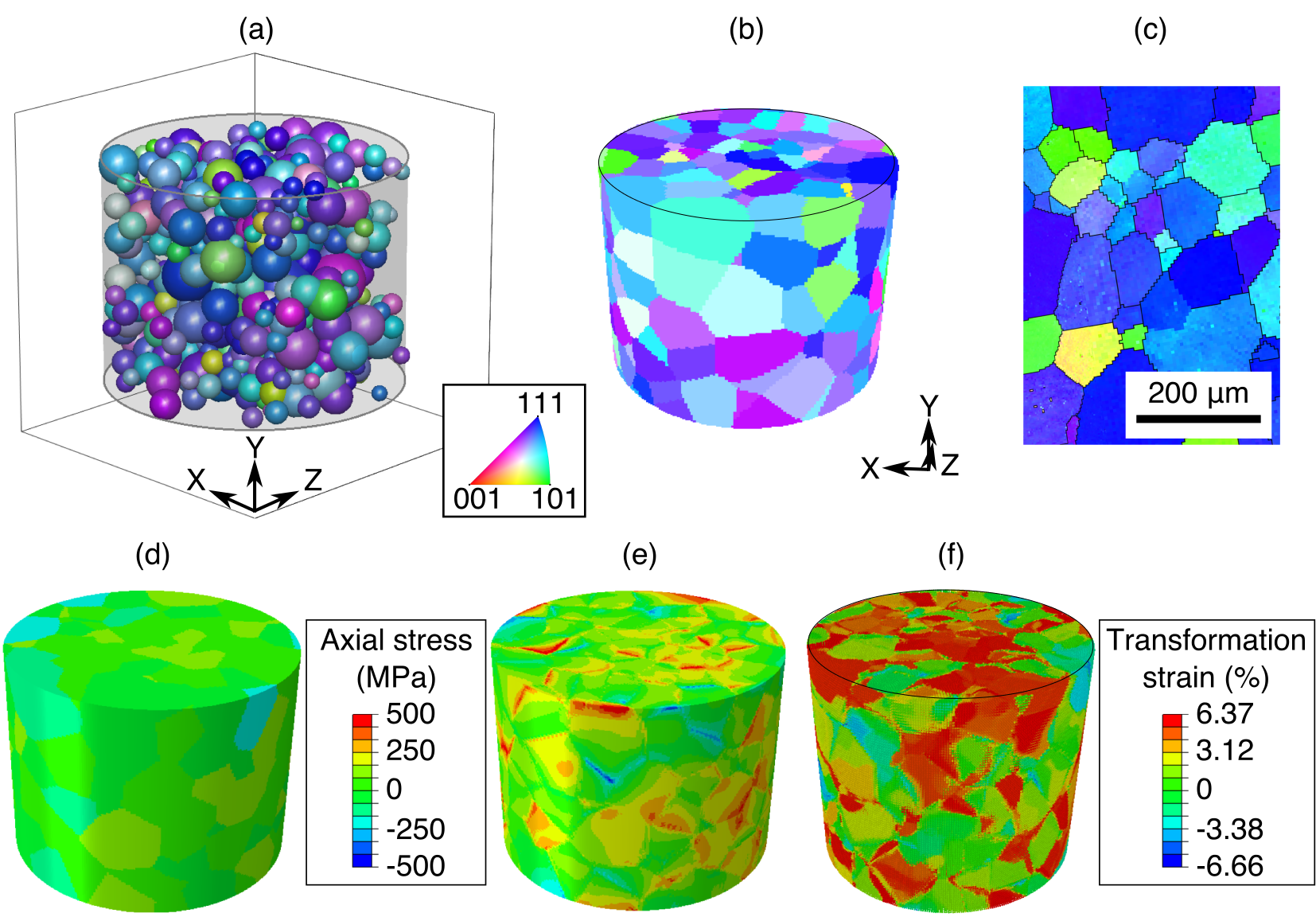}
\caption{Summary of the elastic simulation informed by ff-HEDM. (a) Spatial plot of grain centroids obtained from ff-HEDM at \protect\circled{0} in Cycle 11. Sphere size is proportional to the grain radius and the color corresponds to a cubic inverse pole figure colormap (inset). (b) The Voronoi tessellation obtained from (a). (c) EBSD scan of a slice of the specimen-gage after the ff-HEDM experiment. The majority of grain boundaries is linear, similar to the planar grain boundaries obtained in the Voronoi tessellation. The average grain size in the EBSD scan is \SI{79}{\micro \meter} comparable to \SI{84}{\micro \meter} in the tessellation. The inverse pole figure colormap applies to (a)-(c). (d) Initial (residual) axial stress imposed in an anisotropic elastic simulation of tensile loading with the tessellation. Residual stress tensor is calculated using the lattice strain tensor at \protect\circled{0} from ff-HEDM and B2 stiffness tensor given in section 2.3. (e) Simulated axial stress state at 0.4\% macro strain. The axial stress shows a broad scatter --- from -500 MPa to 500 MPa. The stress colorbar applies to (d) and (e).  (f) Transformation strain from the plate $P_{t}$ predicted to form based on the maximum work criterion (Equation \ref{eq:max_work_plate}). The stress state at 0.4\% macro strain is used for the calculation.}
\label{fig:tess_results}
\end{figure}

Figure \ref{fig:tess_results}(a) shows the grain centroid and crystal orientation data at \circled{0} used to construct the Voronoi tessellation in (b). In a Voronoi tessellation, the cells are convex and the cell boundaries are planar. Figure \ref{fig:tess_results}(c) shows the EBSD scan taken after the ff-HEDM experiment from the cleaved specimen. The grain boundaries in the EBSD scan are predominantly linear, thus they are expected to be planar in 3D. This indicates that the grain shapes generated by the Voronoi tessellation are realistic. The mean grain size is \SI{79}{\micro \meter} in the EBSD image vs. \SI{84}{\micro \meter} in the tessellation. The grain size is determined according to the ASTM linear intercept method \citep{astm_international_astm_2013}.  From the lattice strain state obtained from the ff-HEDM data at \circled{0} and the cubic elastic properties for the B2 phase, residual stresses in each grain are calculated. In the elastic simulation, each grain is assigned the residual stress state at the start of the simulation as shown in Figure \ref{fig:tess_results}(d). The assigned stress state at the start of the simulation is homogeneous at the grain scale and can violate equilibrium at the grain boundaries. However, in the first increment in the simulation, a stress field that satisfies compatibility and equilibrium is obtained. 

The results from the simulation show two remarkable features --- a broad spread in local stress and a resultant substantial heterogeneity in the predicted transformation strain. Figure \ref{fig:tess_results}(e) shows the axial stress distribution at the macro strain of 0.43\%. The mean axial stress is 136 MPa. The standard deviation of 57 MPa, however, is substantial. In fact, there are elements that have stress as high as 500 MPa and as low as -500 MPa. This heterogeneity in stress is not just intergranular, but inside the grains as well. In several instances, the elements with the largest deviation from the mean stress are situated near the grain boundaries.

Such disparity in the stresses is naturally reflected in the martensite microstructure predicted to form. Figure \ref{fig:tess_results}(f) shows the axial transformation strain predicted from the simulated stress state at 0.4\% macro strain. Maximum work criterion for martensite HPV selection, defined in Equation \ref{eq:max_work_plate}, is used to identify the plate likely to activate first. The transformation strain for that plate is calculated. The mean axial transformation strain is 3.98\% with a standard deviation of 2.37\%. The full range of transformation strains calculated at elements spans -6.66\% to 6.37\%.  This result shows a strong heterogeneity in predicted transformation strains, despite a strong \hkl<1 1 1>, \hkl<1 1 0> texture.

\section{Discussion}
\label{sec:discussion}

The grain-scale deformation response in the specimen is highly heterogeneous. This is particularly evident in two results. First, the loading response in Figure \ref{fig:ss_macro}(d) departs from a linear behavior at less than 0.5\% strain. Correspondingly, the visualization of tracked grains in Figure \ref{fig:grain_tracking} indicates that some of the grains started transforming before the onset of nonlinearity in the macro stress-strain response, seen as grains disappearing in loading steps 1 and 2. Furthermore, some grains did not transform at all, even at full load, seen as about 100 grains tracked in Figure \ref{fig:grainnum}(a) at loading step 4, and a 65\% B2 phase volume fraction in Figure \ref{fig:grainnum}(c). Second, the disparity in the mean lattice strains for surface vs. interior grains in Figure \ref{fig:lattice_strains} indicates that various grains were exposed to markedly different stress states right from the start of 11\textsuperscript{th} cycle. Interaction between grains in the specimen, varying magnitude of constraint on the surface grains vs. interior grains, the grain orientation, and interaction between untransformed and transformed regions inside a grain are some of the likely factors causing the heterogeneity in response. We systematically analyze these factors in the subsections below. The discussion first analyzes the ff-HEDM results to obtain specimen-wide statistics relevant to these phenomena. Later we utilize the anisotropic inelastic simulation results to investigate sub-grain scale trends in the heterogeneity of response.

\subsection{Interior Grains Carry Larger Axial Stress than Surface Grains on Cyclic Loading}
\label{sec:discussion_surf_effects}

Figure \ref{fig:lattice_strains_radial} shows that tensile axial lattice strains accumulated within the grains located in the specimen interior (radial distance near \SI{0}{\micro \meter}) vs. much smaller tensile-to-compressive lattice strains within the surface grains (radial distance near \SI{500}{\micro \meter}). The stress-strain curves for cycles 1-10  in Figure \ref{fig:ss_macro}, that gave rise to these residual strains, show both nonlinearities and residual strain accumulation. This result suggests that plastic deformation and stabilized martensite accumulated during the first 10 cycles. The initial B2 phase fraction of 85\% at the start of cycle 11 (Figure \ref{fig:grainnum}(c)) confirms that both plasticity and retained martensite played roles. Based on this information, we propose the following mechanism for the trend in the axial lattice strain at \circled{0} and the grain radial distance.

We hypothesize that the grains with a larger number of neighbors will have a larger axial lattice strain at \circled{0}. Grains in the interior of the specimen are surrounded by neighbor grains on all sides, and thus, on average, have a larger number of neighbors. These interior grains can be considered more constrained than the surface grains, since the grains on the surface can deform unrestricted in at least one direction. The constraint on the interior grains potentially generated a triaxial stress state during the loading in first ten cycles, and the interior grains were not able to plastically deform in the axial direction as much as the surface grains. Additionally, in the interior grains, a variety of slip systems could have activated in order to deform under constraint vs. fewer slip systems in the surface grains. Such a disparity in slip system activation has in fact been observed in crystal plasticity simulations of synthetic microstructures \citep{barbe_intergranular_2001}. This disparate slip activity would result in a larger axial deformation in the surface grains compared to the interior grains. At the end of ten cycles, when the specimen is brought to zero load, the surface grains would experience a compressive strain to maintain compatibility with the neighbors, while the interior grains would be in tension.

\begin{figure}
\centering
\includegraphics[width=5.51in]{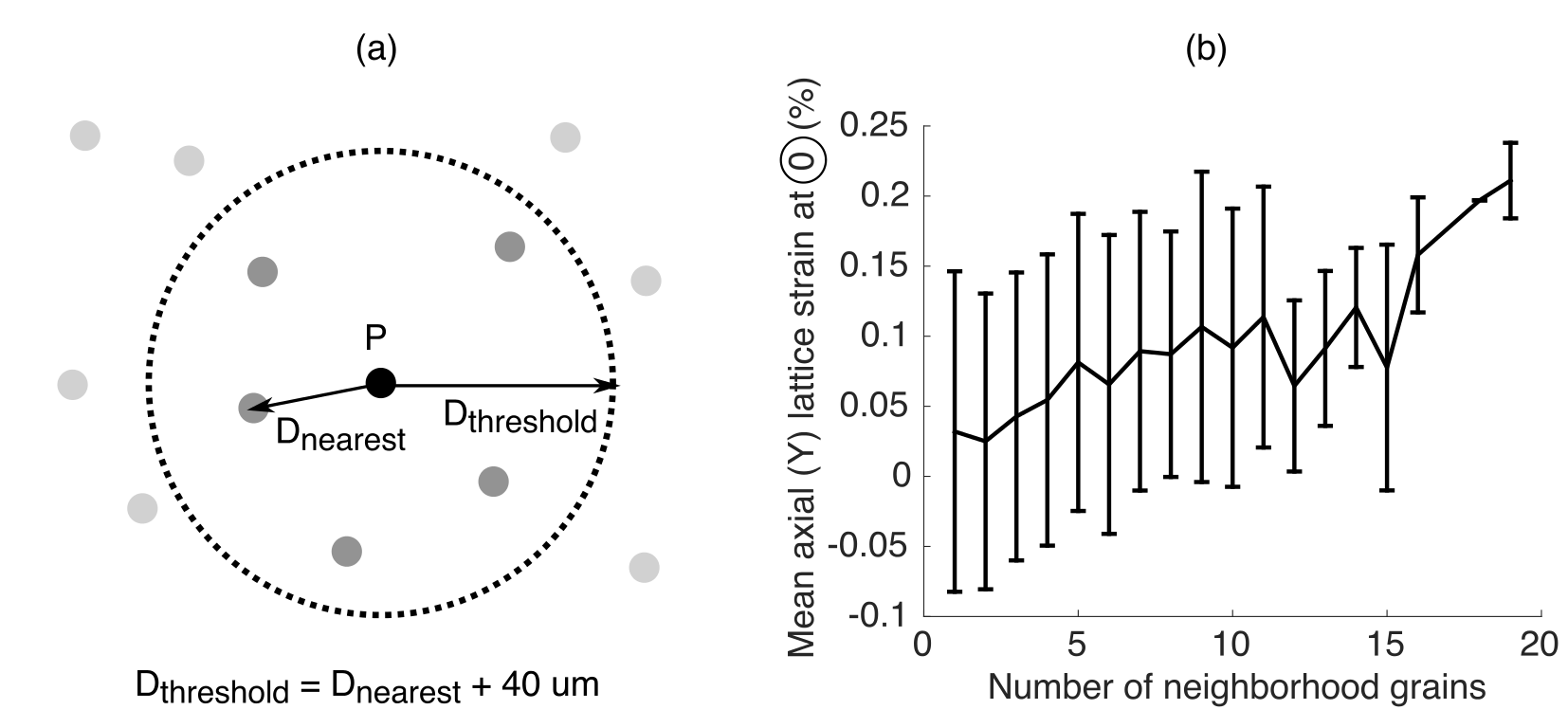}
\caption{The axial lattice strains at \protect\circled{0} correlate with the number of neighbors of a parent grain (P). (a) Schematic of a grain neighborhood. Since ff-HEDM furnishes grain centroid positions, a distance cutoff criterion can be used to count the neighboring grains of a parent grain. The neighbors of P are showing in a darker shade of gray. (b) Axial lattice strain in the grains at \protect\circled{0} as a function of the number of grains in the neighborhood. Parent grains with more neighbors tend to have larger axial lattice strains.}
\label{fig:lattice_strains_neighbors}
\end{figure}

To test this hypothesis, the number of neighbors of each grain need to be counted. Figure \ref{fig:lattice_strains_neighbors}(a) illustrates the algorithm used to count the number of neighbors of a parent grain. For each grain indexed at \circled{0}, the distance of the nearest grain is obtained and a cutoff value \SI{40}{\micro \meter} larger than that distance is set. The cutoff value of \SI{40}{\micro \meter} is chosen to be slightly less that the mean grain radius of \SI{53}{\micro \meter} measured at \circled{0} using ff-HEDM. This choice was made to minimize counting second-nearest neighbors.

A plot of mean axial strain at \circled{0} vs. the number of neighbor grains validates the hypothesis that the grains with a larger number of neighbors have a larger, tensile lattice strain at the start of the 11\textsuperscript{th} cycle. Figure \ref{fig:lattice_strains_neighbors}(b) shows the correlation between the number of neighborhood grains counted using the above mentioned strategy and the mean axial lattice strain at \circled{0} for the parent grain with that many neighbors. Grains with the largest number of neighbors have 0.2\% axial lattice strain vs. 0.03\% for the grains with the least number of neighbors. The trend is accompanied by a large scatter. This is anticipated since the actual neighborhood composition in terms of grain size, orientation, and grain boundary features is expected to be widely varying for the grains with the same number of neighbors.

Overall, it can be concluded that the apparent surface effect observed in the axial strain distribution is a manifestation of the different nature of confinement of the surface grains vs interior grains. While it is well known in plastically deforming materials that a heterogenous stress state exists from the surface to the interior \citep{macherauch_x-ray_1966}, these results furnish unique grain-averaged, specimen-wide statistics of lattice strain heterogeneity in SMAs as a result of combined plasticity and phase transformation. We anticipate seeing a similar influence of the neighbor constraint at the sub-grain scale. 

\subsection{Grains With Similar Initial Axial Strains or Stresses Perform Differently if Their Neighborhoods Substantially Differ}
\label{sec:discussion_intra}

The elastic simulation results in Figure \ref{fig:tess_results}(e) show significant stress heterogeneity at both specimen-scale and grain-scale. Specifically, grain boundary regions tend to show larger deviation from the mean axial stress compared to the grain interiors, and those deviations vary throughout the specimen, ranging from compressive to tensile axial stresses. Such localization of stress deviation is anticipated on the basis that the interaction with the neighbors is strongest in the regions near grain boundaries \citep{barbe_intergranular_2001, wong_stress_2015}. This stress heterogeneity is naturally expected to lead to disparate driving forces for martensite formation and the transformation strains subsequently produced. The results in Figure \ref{fig:tess_results}(f) show a large spread in the predicted transformation strain as anticipated. These results indicate that even in highly textured SMAs, phase transformation is expected to proceed with a significant heterogeneity.

We hypothesize that similarly oriented grains with similar initial residual strain or similar residual stress will perform differently if the neighborhoods are sufficiently different. This hypothesis is based on the heterogeneous stress distribution inside grains in Figure \ref{fig:tess_results}(e). The effect of grain-neighbor interactions is already evident from the heterogeneous deformation introduced during the cyclic loading. Thus it is likely that grains that are initially of similar orientation and strain deform differently in cycle 11 if the grain neighborhoods exert substantially different influences.

To test this hypothesis, we compare the experimental and simulated response of two similarly oriented grains. We consider two distinct cases. First the evolution of experimental strain and simulated stress in two similarly oriented grains with {\em similar normal strains} at \protect\circled{0} is considered. In the next subsection, evolution in the strains and the stresses in two similarly oriented grains that had {\em similar stress states} at \protect\circled{0} are compared.

\begin{figure}
\centering
\includegraphics[width=5.51in]{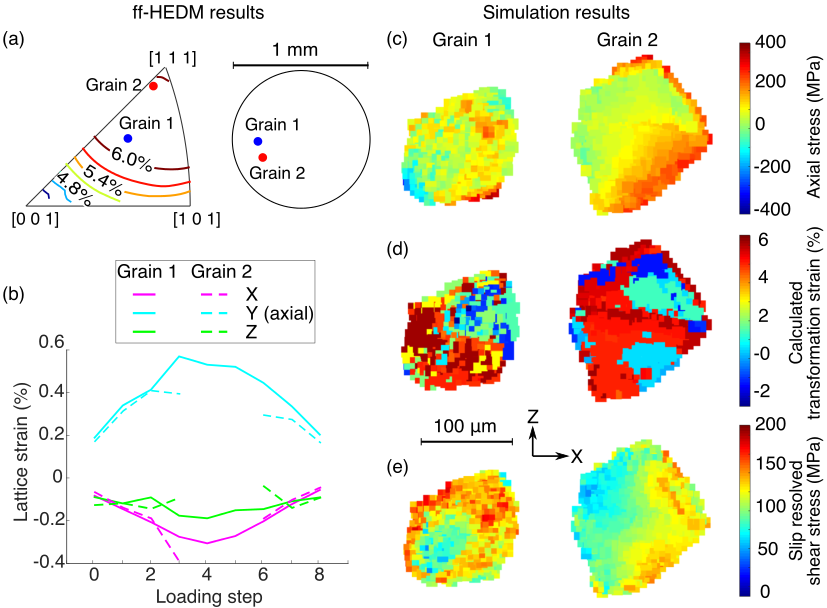}
\caption{Comparison between two grains with similar theoretical transformation strain and initial normal strain state, showing that the grains with similar orientation and residual lattice strain will deform differently if their neighborhoods are substantially different. (a) An inverse pole figure showing the crystal orientations of Grain 1 and Grain 2, and a top view of the specimen showing that the grains are at a similar radial distance. (b) Evolution of the lattice strains during Cycle 11 for Grain 1 and Grain 2 in the ff-HEDM experiment. At loading steps \protect\circled{4} and \protect\circled{5}, Grain 2 disappears; likely due to the grain substantially transforming to B19$^\prime$. (c) Axial stress from the anisotropic elastic simulations at a macro strain of 0.4\%. This macro strain approximately corresponds to \protect\circled{2} in the main cycle of ff-HEDM experiment --- the strain where significant number of grains started transforming. Grain 2 on average has larger axial stress than Grain 1. (d) Using the simulation stress state at 0.4\% macro strain, the B19$^\prime$ habit plane variant (plate) favored to form is determined based on Equation \ref{eq:max_work_plate}. The figure shows the axial transformation strain produced by that plate. On average, Grain 2 would theoretically produce larger transformation strain and it is more likely to transform compared to Grain 1. (e) Resolved shear stress on the slip system favored by the simulation stress state at 0.4\% macro axial strain. On average Grain 1 has larger resolved shear stress vs. Grain 2.}
\label{fig:tale_of_2_grains}
\end{figure}

\subsubsection{Grains with Similar Orientation and Initial Normal Strains}

To test the hypothesis that similarly oriented grains with similar initial lattice strains deform differently if their neighborhoods are substantially different, we consider two grains --- Grain 1 and Grain 2 with such crystal orientations that they are predicted to produce the same transformation strain along the loading direction according to CTM (Figure \ref{fig:tale_of_2_grains}(a)). An equivalent of this condition in plasticity would be to select two grains with the same Schmid factor. The radial distance of the grains from the specimen axis is similar (Figure \ref{fig:tale_of_2_grains}(a)). An additional constraint in selecting these grains is that they had similar \textit{normal} lattice strains at \circled{0} (Figure \ref{fig:tale_of_2_grains}(b)). Independent of granular constraints, these grains should exhibit similar responses to axial loading according to these conditions. However, the initial normal stresses in these grains, calculated using the cubic elastic stiffness for B2 (specified in Section \ref{sec:simulations}), are not similar. The axial stress in Grain 2 is larger than Grain 1 at \circled{0} by approximately 60 MPa, due to different shear strains in the grains at  \circled{0}. The full strain and stress state in Grains 1, 2 is shown in the Supplementary Data  Table S1. This effect propagated through the loading in cycle 11. Figure \ref{fig:tale_of_2_grains}(b) shows that the lattice strains in the two grains evolved differently beyond loading step \circled{2}. Grain 2 is not tracked at step \circled{5}, indicating that it mostly transformed to martensite. The axial (Y) lattice strain evolution in Grain 2 is relaxed in \circled{2}-\circled{3} and again in \circled{6}-\circled{7} from the initial linear loading slope exhibited in \circled{0}-\circled{2} and the corresponding unloading slope in \circled{7}-\circled{8}. The axial lattice strain in Grain 1 monotonically increased until \circled{3} and then monotonically decreased. To investigate the cause behind this disparate behavior between Grain 1 and Grain 2, we compare the simulated stress states in the two grains at the macro strain of 0.4\%, which approximately corresponds to \circled{2} in Figure \ref{fig:ss_macro}(d).

Grain 2 has larger regions with axial stress exceeding 200 MPa, compared to Grain 1 at 0.4\% macro strain, as shown in Figure \ref{fig:tale_of_2_grains}(c). In fact, Grain 1 shows certain regions with a large compressive axial stress, where Grain 2 does not. Grain 1 also exhibits a smaller mean axial stress with a larger spread (44 $\pm$ 72 MPa) compared to Grain 2 (118 $\pm$ 56 MPa). This stress state is reflected in the predicted transformation strains and resolved shear stress on slip systems at \circled{2} shown in Figure \ref{fig:tale_of_2_grains}(d) and (e) respectively. The transformation strains are calculated based on the maximum work criterion previously specified in Equation \ref{eq:max_work_plate}. Grain 2 shows larger regions with a transformation strain greater than 4\% compared to Grain 1. The mean transformation strain in Grain 2 is 4.44\%, compared to 4.17\% in Grain 1. The standard deviation in the transformation strain is 2.14\% for Grain 2 vs. 2.37\% for Grain 1. The mean resolved shear stress on the most efficient slip system ($S_{t}$) in Grain 1 and Grain 2 from the stresses shown in Figure \ref{fig:tale_of_2_grains}(c) are comparable (91 vs. 88 MPa). These results can be used to rationalize the difference in the response in Grain 1 vs. Grain 2.

The higher mean axial stress at 0.4\% macro strain (i.e., \circled{2}) in Grain 2 vs. Grain 1 is certainly the strongest factor favoring phase transformation in Grain 2 vs. Grain 1. However, a contribution from neighborhood interactions is evident from the standard deviations in the simulated axial stress and the predicted transformation strain --- both are larger in Grain 1 vs. Grain 2. Since it is essential to maintain compatibility between grain neighbors during loading, a stronger neighbor interaction would result in a larger stress transfer between the parent grain and the neighbors and encourage the parent grain to deviate more from the macro stress state.

\begin{figure}
\centering
\includegraphics[width=\textwidth]{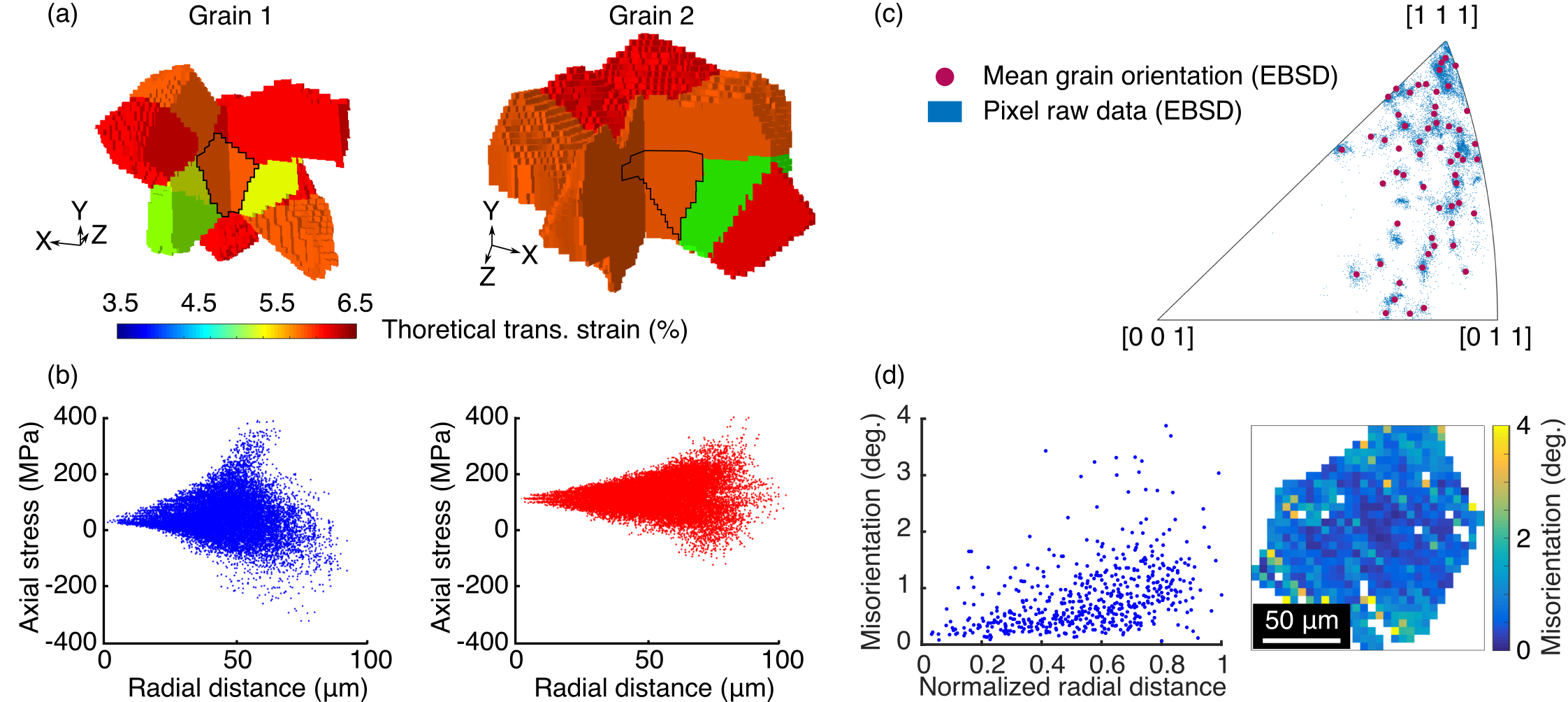}
\caption{Grain 2 has an easy-to-transform grain neighborhood which creates a favorable stress state for transformation. (a) Comparison of the theoretical transformation strains of the nearest neighbors of grains 1 and 2 introduced in Figure \ref{fig:tale_of_2_grains}. Grain 2 has a more homogeneous neighborhood. The figure shows a 3D section of the grain neighborhoods in the tessellation created from ff-HEDM data. (b) Variation of axial stress inside the grains 1 and 2 as a function of the distance from the grain centroid at a macro axial strain of 0.4\% in the elastic simulation. Grain 1 shows a larger scatter, and a lower mean value for the stress compared to Grain 2. (c) EBSD scan of a section of the specimen gage after ff-HEDM test shows a spread in orientations around the mean Grain orientation. (d) Misorientation from the mean grain orientation vs. distance from the grain centroid for one of the grains in the EBSD scan. The misorientation is lower near the centroid and larger in the periphery. The trend however is weak. Grain in (d) is not the same as either of Grain 1 or 2.}
\label{fig:intra_grain_hetero}
\end{figure}

We extend this analysis of neighbor interactions and propose a specific mechanism by which Grain 2 transforms, while Grain 1 does not. We hypothesize that easy-to-transform in-plane neighbors transferred tensile stresses to Grain 2 during loading in cycle 11, and thus promoted phase transformation. The magnitude of stress transferred among the in-plane neighbors is proportional to the axial strain mismatch between the neighbors \citep{li_compact_2001}. Thus an easy-to-transform in-plane neighborhood will transfer tensile stresses to the parent grain and promote transformation, while a difficult-to-transform in-plane neighborhood will transfer compressive stresses, and inhibit the transformation in the parent. This mechanism has been numerically demonstrated to influence the transformation strains generated in NiTi SMA polycrystals during superelastic loading \citep{paranjape_texture_2014}.

We test this hypothesis by comparing the actual nearest neighbors from the tessellation and the intragranular distribution of stresses in Grains 1, 2. Figure \ref{fig:intra_grain_hetero}(a) shows Grains 1, 2 and their nearest neighbors in the tessellation. The grains are colored according to the axial transformation strain predicted based on their orientation. Grain 1 has 14 neighbors with the mean predicted transformation strain of 6.1\%. Grain 2 has 16 neighbors. However its neighborhood is more favorably oriented to transform, with the mean predicted transformation strain of 6.15\%. In particular the in-plane neighbors of Grain 2 show a mean transformation strain of 6.1\% vs. 6.0\% for Grain 1. Additionally, Grain 1 is immediately surrounded by three in-plane neighbors that are rather hard to transform (grains with yellow and green color), where Grain 2 only has one such neighbor. While in-plane neighbors can effectively transfer stresses to the parent grain and promote transformation, the difference in the theoretical transformation strains of the in-plane neighborhoods of Grains 1, 2 is small. Thus these data weakly support the hypothesis that the stress transfer from the in-plane neighbors favored transformation in Grain 2. Overall, Grain 2 is favored to transform compared to Grain 1 based on multiple factors--- higher inherited axial stress after cycling, easier to transform in-plane neighbors which aid transformation by redistributing a tensile stress during loading, and a neighborhood that is more homogeneous and favorably oriented to transform.

Another influence of the different in-plane neighborhoods between Grains 1 and, even through the neighborhood differences are subtle, can be seen in the intragranular stress spread in the two grains. A plot of axial stress in the two grains vs. distance from the respective grain centroid, as shown in Figure \ref{fig:intra_grain_hetero}(b), displays two patterns. First, the spread in the axial stress is larger away from the grain centroid, near the grain boundaries. This underscores the larger influence of grain neighborhood in the periphery of the grain. Second, the axial stress in Grain 1 shows a larger spread compared to Grain 2 --- a consequence of stronger neighbor influence. The stress spread in Grain 1 is large at a relatively smaller radial distance compared to Grain 2. While this could be, in part, due to Grain 1 being smaller than Grain 2, it is likely to be due to stronger neighbor interactions.

The intra-grain stress heterogeneity is reflected in the grain orientation data obtained using EBSD after the ff-HEDM experiment. Figure \ref {fig:intra_grain_hetero}(c) shows the raw pixel data and the mean grain orientations from the EBSD scan of a section of the tension specimen gage. The pixel data shows a spread around the mean grain orientations. The pixel spread or the intra-grain misorientation is likely to be a result of heterogeneous plastic deformation and retained martensite in the grains. The intra-grain stress plots in (b) further suggest that the misorientation is likely from the regions near the grain boundaries. This is indeed seen in the EBSD results. Figure \ref{fig:intra_grain_hetero}(d) shows the intragranular misorientation vs. radial distance from the grain centroid for one of the grains in the EBSD scan. The misorientation is larger and more scattered away from the grain centroid.

These results connect to several reports in the literature of the intra-grain heterogeneous response in inelastically deforming materials. Merzouki et al. show that at around 0.8\% macro strains, the stress concentration sites are predominantly along the grain boundaries in a superelastic simulation of a CuAlBe SMA. Their simulations are informed by the grain structure obtained from EBSD analysis of planar specimen. They further show that numerous martensite plates are activated in individual grains, some of them confined to the regions close to the grain boundaries \citep{merzouki_coupling_2010}. Mika and Dawson observed, using crystal plasticity simulations of deformation in a virtual face centered cubic (FCC) polycrystal, that the deformation gradient deviates more from the mean value as the distance from the grain centroid increases \citep{mika_effects_1998}. Further, they conclude that it is beneficial to use realistic grain shapes in simulations if the aim is to obtain the intragranular deformation heterogeneity during loading. This underscores the benefit of using microstructural data from ff-HEDM to inform the simulation in this study. A similar observation is made by Raabe et al. based on coupled DIC-crystal plasticity study \citep{raabe_micromechanical_2001}. They observed that the grain interaction zones, identified on the basis of inelastic strain localization are strongly clustered around grain boundaries.

This section discussed two sources of deformation heterogeneity --- inherited stress state on cyclic loading and neighbor interaction. While the detailed comparison between Grains 1 and 2 was based on the two grains having similar strains at \circled{0}, they had different axial stresses as a result of cubic elastic parameters of B2 phase. This raises the question --- do grains with a similar initial stress state behave differently due to different neighborhoods? To answer this question, we compare the performance of two grains with similar stresses at \circled{0}.

\subsubsection{Grains with Similar Orientation and Initial Stress}

\begin{figure}
\centering
\includegraphics[width=\textwidth]{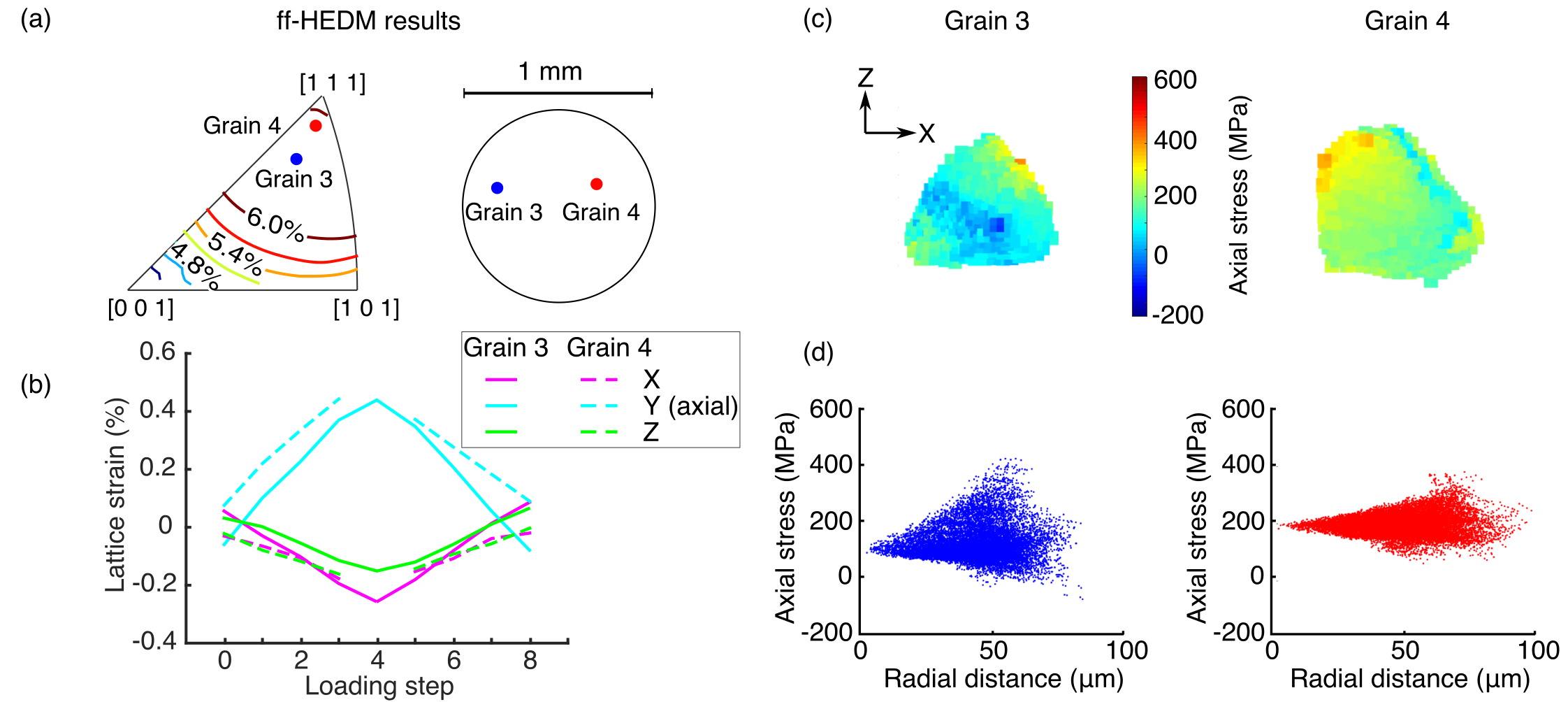}
\caption{Comparison between grains 3, 4 with similar initial stress state and orientation shows that Grain 4 developed a higher mean axial stress with lower scatter vs. Grain 3 due to neighbor interactions and thus transformed. (a) Crystal orientation and grain centroid. (b) Evolution of lattice strains. Grain 4 is not tracked at \protect\circled{4}, potentially due to transformation to B19$^\prime$. (c) Axial strain from the anisotropic elastic simulation at 0.4\% macro strain. Overall, axial stress is higher in Grain 4. (d) Axial stress vs. distance from the grain centroid at 0.4\% strain in the simulation. Grain 3 shows a larger heterogeneity and a smaller mean stress value compared to Grain 4.}
\label{fig:tale_of_other_2_grains}
\end{figure}

To test the hypothesis that, similarly oriented grains with similar stress states at \circled{0} will deform differently in cycle 11, if their neighborhoods are substantially different, we compare the deformation in Grain 3 and Grain 4, that are similarly oriented and had a similar stress state at \circled{0}. The orientation and radial position of Grains 3, 4 are shown in Figure 11(a), and the complete stress and strain states are provided in the Supplementary Data Table S1. During loading, Grain 4 transformed, while Grain 3 did not as shown by the break in the tracked lattice strains in Figure \ref{fig:tale_of_other_2_grains}(b), despite the two grains having similar predicted mean transformation strains --- 6.27\% for Grain 3 vs. 6.07\% for Grain 4 as shown in Figure \ref{fig:tale_of_other_2_grains}(a). Grain 4 developed a higher mean axial stress (Figure \ref{fig:tale_of_other_2_grains}(c)) with a tighter spread (Figure \ref{fig:tale_of_other_2_grains}(d)) vs. a lower mean stress and a wider stress spread in Grain 3, despite them having similar initial stresses. The axial stress in both grains is approximately 98 MPa at \circled{0} and the values of all other stress components compared between the two grains are within 30 MPa (Supplementary Data Table S1). Their initial normal strains however, are different (Figure \ref{fig:tale_of_other_2_grains}(b)). The simulated axial stress at 0.4\% macro strain is 131 MPa in Grain 3  with 61 MPa standard deviation. The mean/standard deviation in the axial stress in Grain 4 at the same macro strain is 197 $\pm$ 35 MPa. The larger axial stress on loading in Grain 4 vs. Grain 3, despite them having similar stresses at \circled{0}, indicates that Grain 4 developed a favorable stress state for phase transformation due to its neighborhood. This observation validates the hypothesis that similarly oriented grains with a similar initial stress state can deform differently due to different neighbor influences. While this result is similar to the observations for Grains 1 and 2, the disparity in the stress states for Grains 3, 4 is definitively from the neighbor interactions.

This section showed that the neighbor constraint generates a heterogeneous stress state, particularly near the grain boundaries. This heterogeneity is in addition to the heterogeneous residual stresses generated on cyclic loading. The grains that transform, are likely to have higher and more homogeneous axial stresses compared to their constrained, non-tranforming counterparts. Based on this constraint mechanism, we can expect to observe certain trends in the grain-scale stresses, stress heterogeneity, and grain rotations as a function of the size and position of the grains. Since phase transformation is strongly dependent on crystal orientation, we should expect to see some trends based on grain orientations. These trends are inspected in the next three sections.

\subsection{Surface Grains Show a Larger Axial Stress Spread}
\label{sec:trends_grain_stress_hetero}

We hypothesize that the surface grains will show a larger stress spread compared to interior grains at a fixed macro strain in the simulation. This hypothesis is motivated by the observation that the interior grains are uniformly constrained, while the surface grains are unconstrained in at least one direction, and may have varying degree of constraint due to neighbors in other directions. In a grain, interaction with neighbors creates a stress spread, particularly near the grain boundaries. Thus, we would expect the stresses in the grains near the surface, with their relatively low constraint, to evolve nearly same as the macroscopically imposed stresses. However, parts of the the surface grains that are away from the free surface and adjacent to the interior grains would deviate substantially from the macro stress state. Interior grains, on the other hand, would be unlikely to keep up with the macro stresses and uniformly deviate from the macro stress state. The grains in the interior inherit a tensile strain at \circled{0}. Hence the stress state in the interior grains is biased towards large tensile values. Surface grains do not have such bias since they inherit smaller axial strains at \circled{0}. 

\begin{figure}
\centering
\includegraphics[width=5.51in]{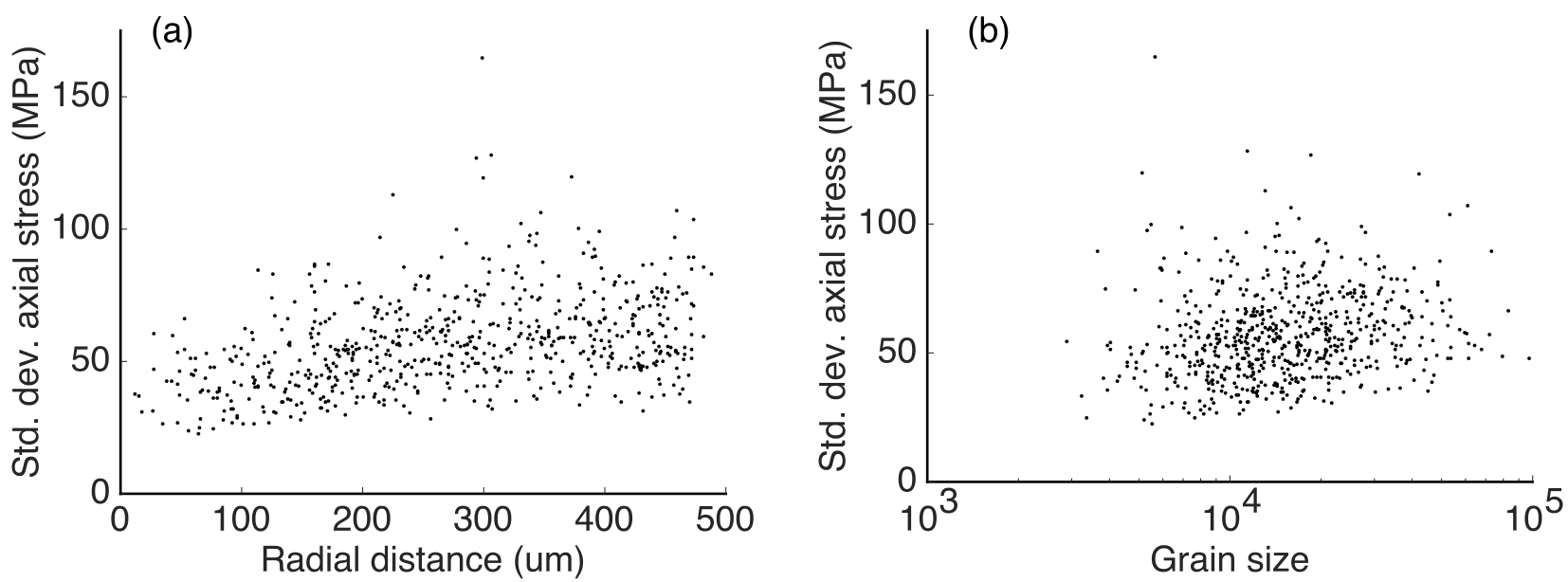}
\caption{Statistically, simulated grain-scale axial stress heterogeneity depends on whether the grain is in the specimen interior or near the surface, but does not depend on the grain size. (a) Standard deviation in the axial stress in the grains at 0.4\% macro strain in the anisotropic elastic simulation shows a weak trend with the grain radial distance. (b) Standard deviation in the grain axial stress plotted as a function of grain size (measured as the number of finite elements) does not show a significant trend, however larger grains have a larger lower bound on the stress spread vs. smaller grains.}
\label{fig:intra_grain_hetero_2}
\end{figure}

To validate this hypothesis, we inspect the trend in the simulated axial stress spread vs. radial distance. Figure \ref{fig:intra_grain_hetero_2}(a) shows the standard deviation in the intra-grain stress vs. radial distance of grain centroid from the specimen axis. Surface grains show a larger stress spread, supporting the hypothesis. However the trend is weak. The reason behind the weak trend could be that the surface grains in this case have only one free surface and in other directions they are surrounded by very different neighborhoods. This diversity in neighborhoods contributes a scatter to the stress spread statistics.

Extending the observation above to grain sizes, we expect to see larger stress spreads in larger grains. This hypothesis is inspired by the Saint-Venant's principle. Every grain can be thought to compose of a core region that is relatively unaffected by neighbor interaction and a shell region that is strongly influenced by the neighbors. In smaller grains, it can be expected that the neighbor influence is felt throughout.

We extract grain-scale stress spread information and visualize it as a function of the size of the grain, to test if larger grains show larger stress spreads. Figure \ref{fig:intra_grain_hetero_2}(b) shows the standard deviation in the simulated axial stress in the grains approximately at \circled{2} (i.e., at 0.4\% macro strain) vs. grain size. The grain size is expressed in terms of the number of elements in the finite element mesh for that grain and its plotted on a log scale. As such there is no trend and the data do not support our hypothesis. However, we can say that the lower bound on the stress spread for larger grain is larger than the lower bound for smaller grains. This result is similar to the lack of particular trend in strain spread vs. grain size in the empirical work of Kimiecik et al. on planar NiTi specimen \citep{kimiecik_grain_2015}. The reason for the lack of trend could be that the grain size effect is weaker compared to the combined effect of grain orientation, position, and inherited stress state at \circled{0}. Due to these effects, grains with similar sizes show a broad variation in stress states. The role of grain orientation in particular on the heterogeneous response is worth investigating further since the phase transformation response is highly orientation dependent. Subtle differences in the stress state of similarly oriented grains can cause activation of very different martensite plates and result in a lower than ideal level of transformation strains. In the next section, we discuss the role of grain orientation and any disparity in the deformation response of similarly oriented grains. These issues are relevant in determining the suitability of modeling frameworks that treat similarly oriented grains as equivalent, irrespective of their neighborhoods.

\subsection{Similarly Oriented Grains Show Dissimilar Lattice Strain Evolution}
\label{sec:trends_sim_orientation}

\begin{figure}
\centering
\includegraphics[width=3.54in]{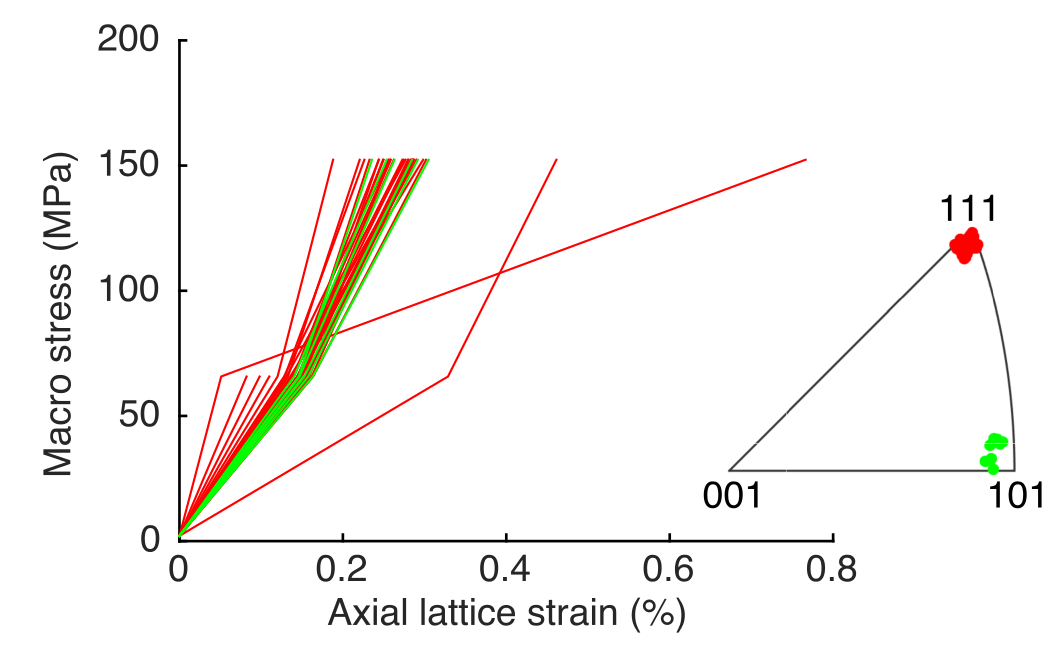}
\caption{The axial lattice strain in similarly oriented grains evolves heterogeneously. Plot of macro stress vs. grain-averaged lattice strain for two sets of grains --- Those with \hkl<1 1 1> axis oriented along the loading direction and those with \hkl<1 1 0> direction oriented along loading. All quantities are for the axial (Y) direction. Difference between the lattice strain at the current loading step and the initial loading step \protect\circled{0} is used to exclude the bias due to heterogeneous strains at the start of the main cycle. The lattice strain in some grains hardens right at loading step \protect\circled{1}, suggesting onset of inelastic deformation. Spread in the stress/strain response of similarly oriented grains attests to the influence of different neighbor interaction.}
\label{fig:latt_strain_macro_stress}
\end{figure}

We anticipate that similarly oriented grains will show different strain states at a fixed macro stress, due to their disparate neighborhoods. Figure \ref{fig:latt_strain_macro_stress} shows the axial macro stress (measured by the load cell) as a function of ff-HEDM axial lattice strain in the grains belonging to two groups. First group consists of grains with \hkl<1 1 1> crystal direction oriented along the loading axis with approximately \SI{10}{\degree} tolerance and the second group consists of \hkl<1 1 0> oriented along the loading axis with the same tolerance. First, the grains show disparate macro stress-lattice strain response. Some grains disappear from tracking below 100 MPa macro stress. Second, the axial lattice strain relaxes, i.e., the macro stress-lattice strain response hardens for most of the grains at the first loading step. These empirical results show that the grain response is heterogeneous even for similarly oriented grains and transformation in some grains starts at a very low stress. The first phenomenon is a consequence of the neighbor interaction, while the second phenomenon can be a consequence of the combined effect of neighbor interaction, the inherited stress state at \circled{0} and any retained martensite present at \circled{0}.

This result, coupled with the observations of heterogeneous stress inside the grains due to varying neighborhoods, discussed in prior section suggest that full-field, micromechanical modeling technique are more suitable to capture the local deformation response in SMAs compared to models with self-consistent or homogenization schemes. While homogenized models provide a computationally efficient means to transition from the grain-scale to the macroscopic response, factors such as grain neighborhood effects are not captured by such methods \citep{lagoudas_shape_2006, gao_simplified_2002}. Full-field models, typically implemented in the finite element frameworks can utilize the microstructural information obtained from ff-HEDM experiments to generate realistic virtual microstructures and explicitly model grain neighbor interactions. Such efforts have been utilized by the plasticity community, specifically to understand grain fragmentation and texture evolution during inelastic deformation \citep{miller_understanding_2014}. SMA community can benefit from similar combined 3D experiment-modeling studies. These efforts are more likely to capture the degradation in superelastic performance and deformation localization due to specific arrangements of grains and their interactions; an ability that could be crucial to understand the functional and structural fatigue in SMAs. This analysis also demonstrates the limited ability of powder diffraction analyses to be sensitive to heterogeneities amongst similarly oriented grains. In a powder experiment, the grains that transform would just stop contributing to the average \hkl<1 1 0> and \hkl<1 1 1> responses, though that would not necessarily result in a dramatic change in the macro-stress vs. lattice strain response, even though significant inelastic deformation has occurred in many grains.

Since, grains within approximately \SI{10}{\degree} of \hkl<1 1 1> and \hkl<1 1 0> are selected for this analysis. It is likely that a fraction of the spread in the stress-strain plots in Figure \ref{fig:latt_strain_macro_stress} is due to that small orientation spread. The two grains with unusually large lattice strains are potentially due to erroneous grain tracking in the ff-HEDM analysis. The analysis in this and previous subsections focussed on the effect of neighborhood interactions on similarly oriented grains. However, inherently the phase transformation response is dependent on the crystal orientation of the grains. In the next section, the consequences of orientation dependence are briefly analyzed.

\subsection{Grain Orientation Influences the Onset of Phase Transformation}
\label{sec:trends_orientation}

\begin{figure}
\centering
\includegraphics[width=5.51in]{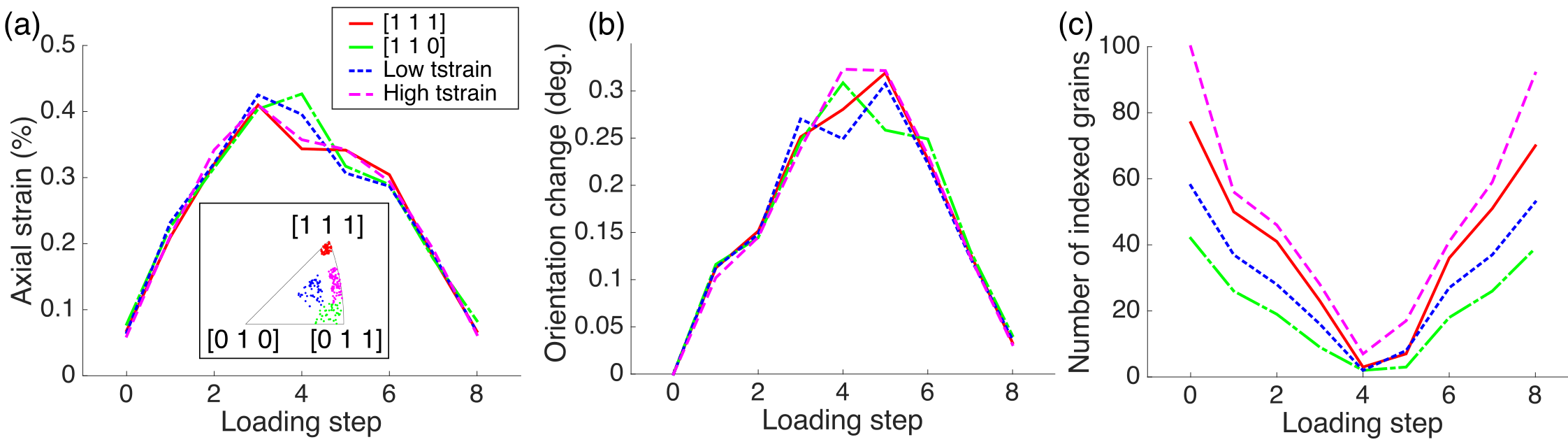}
\caption{Despite neighbor interactions, the influence of crystal orientation on transformation kinetics is visible in the ff-HEDM data. The evolution of (a) axial strain, (b) grain rotation, and (c) the number of B2 grains tracked for four orientation groups. The grain groups are identified with different colors as shown in the inverse pole figure inset in (a). The same color key applies to all subfigures. (a) Grains with \hkl<1 1 1> directions aligned with the loading axis and grains with the largest theoretical transformation strain (tstrain) for B2 $\rightarrow$ B19$^\prime$ transformation tend to have lower axial lattice strain at loading steps 4 and 5 (peak load). (b) \hkl<1 1 1> and {\em high tstrain} groups tend to rotate more near peak load. (c) The number of \hkl<1 1 1> and {\em high tstrain} grains decreases the fastest during loading. These observations can be rationalized on the basis that \hkl<1 1 1> oriented and {\em high tstrain} grains are more likely to transform compared to other two groups.}
\label{fig:orientation_dep}
\end{figure}

We hypothesize that statistically, the grain orientation will be the dominant factor in determining the onset of phase transformation.  To test this hypothesis, we compare the evolution in the averaged ff-HEDM response of grains belonging to four orientation groups, as shown in Figure \ref{fig:orientation_dep}. The orientations of the grains belonging to these four groups are shown in Figure \ref{fig:orientation_dep}(a). The first two groups, \hkl<1 1 1> and \hkl<1 1 0> consist of grains which have \hkl<1 1 1> and \hkl<1 1 0> crystal axis respectively oriented along the loading direction. The {\em Low tstrain} and {\em High tstrain} groups consist of grains that are expected to exhibit lowest and highest theoretical transformation strains respectively among the indexed grains. The transformation strains in this case are calculated based on the grain orientations only. 

The evolution of axial lattice strain vs. loading step, shown in Figure \ref{fig:orientation_dep}(a) reveals that the axial lattice strains evolve similarly for the four orientation groups at the beginning of loading (i.e., up to loading step 3) and at the end of unloading (i.e., after loading step 6). However near the peak stress, \hkl<1 1 1> and {\em High tstrain} groups exhibit lower lattice strains compared to the other two groups. This result validates the hypothesis that \hkl<1 1 1> and {\em High tstrain} grains are easy to transform and thus they partially transformed earlier. This transformation resulted in the stress relaxation in the B2 regions of the grains that are indexed in the ff-HEDM results. The differences in the lattice strains at  loading step 4 between easy-to-transform grains and hard-to-transform grains (\hkl<1 1 0> and {\em Low tstrain} groups) is 0.06\%. The difference can contribute to approximately 40 MPa difference in stresses, based on the B2 elastic constants given in Section \ref{sec:simulations}. It is most likely to be a contribution of the partial transformation-induced stress relaxation, together with the effect of the difference in the elastic stiffness of the grain groups. While the grains in all four groups together are clustered in the right half of the inverse pole figure (Figure \ref{fig:orientation_dep}(a)), the Young's modulus in those grains (along Y direction) can be different by as much as 15 GPa.

Another consequence of the interaction between transformed and untransformed grain regions in various orientation groups is that the grains reversibly rotate during loading/unloading. Figure \ref{fig:orientation_dep}(b) shows the average grain rotation for the grains in the four groups. Grains in the \hkl<1 1 1> and {\em Low tstrain} groups show the largest rotations at the peak stress. This further supports the argument in the previous paragraph that the grains in these two groups transformed to a larger extent compared to those in \hkl<1 1 0> and {\em Low tstrain} groups. The compatibility requirement between transformed and the untransformed parts of the grain is the cause behind the grain rotations. This effect connects to the observations of Berveiller et al., who observed rotations between \SIrange{0.3}{0.5}{\degree} at 0.6\% strain in a superelastic CuAlBe SMA using 3DXRD \citep{berveiller_situ_2011}. Additionally, a contribution from the elastic interaction between grains to the reversible rotation is likely. The B2 grain rotation due to phase transformation is reversible. This is in contrast to the irreversible grain rotation during plastic deformation, e.g., in Cu \citep{pokharel_-situ_2015}. While the results in the last two paragraphs show that the grains in the \hkl<1 1 1> and {\em High tstrain} groups transform more readily and to a larger extent compared to the grains in the other two groups, a well-established result for the B2 $\rightarrow$ B19$^\prime$ transformation in NiTi SMAs, an additional confirmation of that behavior is seen in terms of the tracked grain numbers.

Figure \ref{fig:orientation_dep}(c) shows the evolution in the tracked grain numbers vs. loading step for the four grain groups. As expected the grains in \hkl<1 1 1> and {\em High tstrain} groups diminish in number more rapidly that the other two groups due to the grains in those groups transforming to B19$^\prime$ relatively easily. Thus the influence of grain orientation on the onset of phase transformation is visible in the ff-HEDM results.

\subsection{B2 Grain Rotation is Influenced by the Extent of Phase Transformation }
\label{sec:trends_transformation_fraction_rotation}

\begin{figure}
\centering
\includegraphics[width=5.51in]{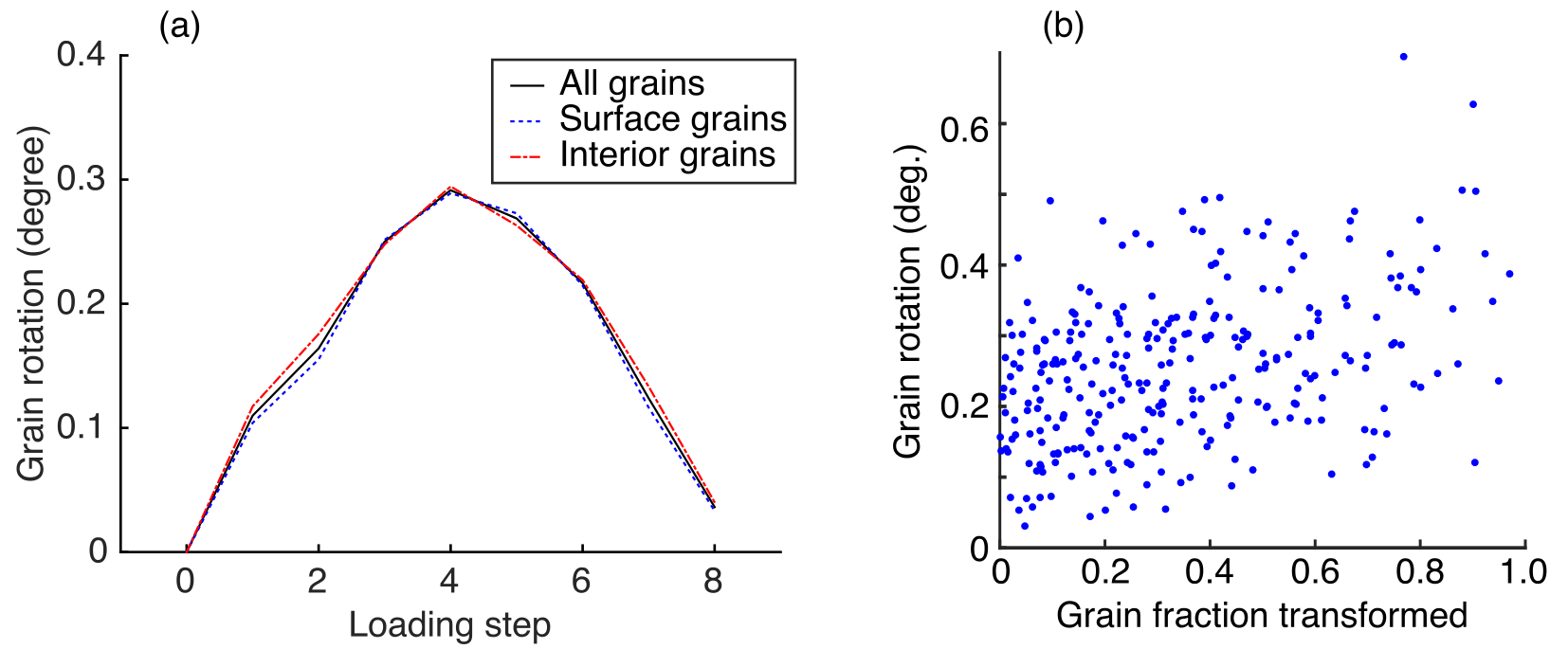}
\caption{Grain rotation during Cycle 11 does not depend on the grain position, but shows a trend with the extent of transformation. (a) Evolution of the change in crystal orientation of all tracked grains. The grains rotate as the load increases and return to the original orientations with some remnant rotation. (b) Maximum change in the orientation of the grains shows a weak correlation with the extent of phase transformation in the grain. This plot is obtained by tracking the rotation and volume of B2 grains across all loading steps. A larger reduction in grain volume implies that a larger fraction of the grain transformed to B19$^\prime$. The compatibility requirement between remnant B2 and transformed B19$^\prime$ region of the grain can require a rotation of the B2 grain.}
\label{fig:grain_rotation_size_change}
\end{figure}

We hypothesize that the rotation in a B2 grain will be influenced by the extent of phase transformation in the grain. This hypothesis is motivated by the intuition that the B2 grains reversibly rotate during phase transformation to maintain compatibility with their transformed regions that have generated large deformations. To test this hypothesis, it is necessary to isolate the influence of other factors such as the grain position on the grain rotation. Separate grain rotation data for specimen surface and interior grains shows that the grain rotation is not influenced by the position of the grain. Figure \ref{fig:grain_rotation_size_change}(a) shows the change in grain orientation averaged over surface and interior grains separately as a function of the loading step. Surface vs. interior grains, on average, rotate by the same amount. This allows us to quantify the effect of interaction between transformed and untransformed parts of the grain on grain rotation. While it is not possible for this experiment to quantify the B2 and B19$^\prime$ volume fraction in each grain using the ff-HEDM technique, we can estimate the extent of transformation in each grain by tracking the B2 grain volume. A grain transformed to a larger extent will exhibit a larger reduction in volume. Figure \ref{fig:grain_rotation_size_change}(b) shows the maximum grain rotation as a function of the extent of transformation. We observe a rather weak trend between the extent of transformation and the grain rotation. Overall the lower bound on the rotation for the grains with a large change in volume is higher than those grains with a smaller change in volume. There could be several reasons behind the weak trend. The grain rotation is likely to be influenced by the neighbor interaction, any plastic deformation in the grains and varying elastic stiffness from grain to grain. In fact, the residual rotation at the end of the 11\textsuperscript{th} cycle in Figure \ref{fig:grain_rotation_size_change}(a) suggests the presence of plastic deformation, at least in some grains. This analysis tacitly assumes that the grain rotations due to elastic deformation alone are comparable for the grains in this aggregate.

\subsection{Challenges and Opportunities for Further Development}
\label{sec:future_develop}

This work demonstrated that ff-HEDM and microstructural modeling together with supporting data such as local/global strains from DIC provide a robust and in fact, essential toolbox to characterize heterogeneous deformation at the grain scale in SMAs. The state-of-the-art in this approach faces certain challenges and there are opportunities for further development.

The analysis in this work focussed on B2 grains and the evolution of the B19$^\prime$ microstructure is not discussed. The reason being that the ff-HEDM technique relies on the position and intensity of spots in the diffraction pattern to obtain grain-scale orientations and strains. The spots can be obtained if the crystallites in the specimen are of at least $\SI{}{\micro \meter}$ size. When the crystallites are smaller, e.g. nm-sized B2 grains or nm-sized twinned B19$^\prime$ plates, the diffraction pattern becomes smeared out, appearing to consist of continuous rings, rather than spots and it is not possible to resolve individual crystallites in the ff-HEDM analysis. It is however possible to analyze the rings in the diffraction pattern corresponding to B19$^\prime$ using powder diffraction analysis techniques and obtain the texture of B19$^\prime$ \citep{hasan_hard_2008}. A more robust solution to obtain spatial variation in the stress-induced B19$^\prime$ microstructure would be to use an iterative forward-modeling approach. In such an approach, the likely B19$^\prime$ microstructure induced in each grain will be obtained using a micro-mechanical or crystallographic modeling framework. The diffraction pattern from such a microstructure will be simulated and compared with the actual diffraction pattern. All likely martensite microstructures will be tested in this manner, until a reasonable match between the actual and simulated diffraction patterns is obtained. As an example, Pagan and Miller have developed a forward modeling approach for slip activity prediction in plastically deforming materials \citep{pagan_connecting_2014,pagan_determining_2016}. We envision following a similar approach for phase transformation, since all B19$^\prime$ orientations that can form in a B2 grain under stress can be calculated using the crystallographic theory of phase transformation \citep{bhattacharya_microstructure_2003}.

This work utilized a Voronoi tessellation scheme to obtain a realistic microstructure for the simulations from the ff-HEDM results. While the grain shapes obtained in the tessellation compare well to the EBSD scan of the specimen, certain other characterization techniques, e.g. near-field HEDM (nf-HEDM) can be utilized to obtain the exact grain morphology \citep{pokharel_-situ_2015}. Similarly, we assigned a uniform orientation to all elements inside a grain in the simulation. A technique like nf-HEDM can furnish intragranular orientation spread that can be incorporated in the simulations, making their results even more realistic. nf-HEDM however does not provide local lattice strain information.

We utilized grain tracking between loading steps to obtain the evolution in deformation. Disappearance of a grain from tracking is seen as a consequence of transformation to B19$^\prime$ phase, which is not indexed. In certain cases, it is likely that the grain plastically deformed substantially, drifted out of the analyzed gage section or fragmented into multiple domains smaller than the approximately \SI{10}{\micro \meter} resolution of this analysis and thus was excluded from tracking. In our analyses consisting of Grains 1 to 4, we specifically selected grains from the middle 70\%  of the gage. Additionally we realigned the specimen at each loading step so that the same region was analyzed during each loading step. Hence the specimen drift is less likely to be a factor adversely affecting this analysis.

We used elastic parameters in the simulation from literature. While it is possible to calibrate the cubic elastic constants from ff-HEDM data \citep{efstathiou_method_2010}, it is not the focus of this study. Further, the presence of inelastic deformation at very low stresses as seen in the grains disappearing early in loading in Figure \ref{fig:grain_tracking} suggests that coupled phase transformation-plasticity simulations would be more appropriate in this ff-HEDM/modeling study.

The ability of this HEDM-modeling approach to obtain material properties (e.g., stiffness tensor), microstructure, and surface grain stresses is likely to be beneficial for optimizing NiTi SMA wire production techniques. SMA wires for example are stress cycled to introduce a compressive strain on the surface \citep{dooley_shape_2004,dooley_method_2012, schaffer_method_2013}. The compressive strain reduces the tendency of micro-cracks to open, thus improving fatigue life of the SMA specimen. A modification of the low-amplitude loading, as performed in this work could potentially be more beneficial in introducing moderate compressive stresses on the surface than the conventional methods. A low-amplitude cycling regime is less likely to introduce large plastic deformation and thus larger flaws on the surface \citep{gupta_high_2015,pelton_situ_2015}.

\section{Conclusions}
\label{sec:conclusions}

This work gives new insight about the effects of granular constraints, the heterogeneity in deformation between specimen surface and specimen interior grains on cyclic loading, and the heterogeneity of deformation in the grain interior vs. grain periphery, by furnishing unique experimental data and reporting on the underlying physics leading to the observed phenomena. Microstructural and deformation data were experimentally measured using far-field high energy diffraction microscopy (ff-HEDM), a non destructive characterization technique that furnishes statistics of grain-averaged crystal orientation and lattice strain, and grain volume. Anisotropic, elastic simulations were performed to augment the experimental deformation data to the subgrain-scale. The experiment informed the simulations in terms of the orientation, position, and residual stresses in the grains, and the austenite lattice parameters. We reported six phenomena related to the heterogeneous deformation in NiTi SMAs.

\begin{enumerate}
\item On cyclic loading, deformation heterogeneity developed between surface grains and interior grains. The interior grains showed larger tensile lattice strains along the loading direction vs. smaller lattice strains in the surface grains. We hypothesized that the surface-interior deformation disparity is a consequence of the interaction between grain neighbors. During the cyclic loading, interior grains which tend to have more neighbors, are more constrained than the surface grains. Those grains with more neighbors do not inelastically deform as much as  the surface grains with fewer neighbors, and thus develop a tensile axial strain state at the end of cycling. An experimental correlation between the residual axial lattice strain and the number of neighboring grains supported our hypothesis.
\item Microstructural simulations informed by the experimental data on grain-scale residual strains, together with experimental B2 grain tracking data showed that the progress of phase transformation in similarly oriented grains is determined by the neighborhoods of the grains and the residual strain state. A larger residual axial stress and a more homogeneous and easy-to-transform neighborhood promotes phase transformation in the parent grain. A heterogeneous and difficult-to-transform neighborhood leads to a lower and more heterogeneous axial stress state in the parent grain, which suppresses phase transformation.
\item Microstructural simulations revealed that specimen surface grains show a larger stress scatter on loading compared to interior grains. Interior grains are constrained all around, thus the stress state in such grains deviates from the macro stress uniformly. The stresses in the unconstrained regions of the surface grains evolve nearly the same as the macro stress, but parts of the surface grains adjacent to neighbors are constrained and thus deviate from the macro stress state, leading to a larger stress spread in the surface grains.
\item Similarly oriented grains showed a spread in the experimentally measured axial lattice strains during loading, even at low stresses, due to the effect of neighbor interactions. Thus, full-field, microstructural models may be more appropriate in simulating the local constitutive response even at smaller stresses.
\item Despite the above evidence for neighbor interaction influencing grain-scale constitutive response in SMAs, phase transformation is still orientation dependent. This is seen in the distinct axial strain evolution, grain rotations, and grain number changes in the grains clustered according to their orientations (e.g., \hkl[1 1 1] and \hkl[1 1 0]) in the experimental data.
\item Experimentally measured grain rotations showed that the magnitude of the rotation in a grain weakly trends with the extent of transformation in the grain. This effect is the manifestation of the compatibility constraint between the untransformed and the transformed regions of a B2 grain.
\end{enumerate}

This work extends the existing paradigm of integrated modeling/experiments to study 3D, grain-scale phenomena in SMAs. This work concretely demonstrates that ff-HEDM can inform microstructural simulations of phase transformation in terms of the initial grain structure, 3D residual strain state, and lattice parameters. The experimentally measured lattice strains during loading can validate models of phase transformation. This combined effort provides statistics about 3D deformation phenomena that are not accessible to other surface-based techniques.

\section*{Acknowledgement}
HMP, PPP, and LCB acknowledge the financial support from Department of Energy, Basic Energy Sciences (grant no. DE-SC0010594). APS acknowledges funding from NSF-Career award no. 1454668. Electron microscopy work reported in this article was performed at NUANCE and OMM Facilities (funded by NSF DMR-1121262) at Northwestern University. This research used resources of the Advanced Photon Source, a U.S. Department of Energy (DOE) Office of Science User Facility operated for the DOE Office of Science by Argonne National Laboratory under Contract No. DE-AC02-06CH11357. The ff-HEDM experiments were performed at 1-ID-E beamline at the Advanced Photon Source. This work used the Extreme Science and Engineering Discovery Environment (XSEDE), which is supported by National Science Foundation grant number ACI-1053575. Technical support from Prof. B. Kappes (Colorado School of Mines) is acknowledged.
\section*{References}
\bibliographystyle{model2-names.bst}\biboptions{authoryear}
\bibliography{library_constraints_2016_v3}

\end{document}